%% file: main.tex
\documentclass[11pt]{article}
\usepackage{graphicx} 
\usepackage{setspace}
\usepackage{amssymb}
\usepackage{amsfonts}
\usepackage{amsmath}
\usepackage{comment}
\usepackage[comma]{natbib}
\usepackage{marginnote}
\usepackage{enumitem}
\usepackage{xcolor}
\usepackage{subcaption} 
\usepackage{url}
\usepackage{array}
\usepackage{enumitem}
\usepackage{booktabs}
\usepackage[margin=1.25in]{geometry}
\usepackage[colorlinks=true,linkcolor = blue, citecolor = black,  urlcolor = black]{hyperref}

\usepackage{amsthm}
\usepackage{tikz}
\usetikzlibrary{trees}

\newtheorem{lemma}{Lemma}
\newtheorem{proposition}{Proposition}

\title{Reactive Marketing and the Co-Production of (In)Authenticity}

\date{} 
\begin{document}
\begin{titlepage}

\noindent
\textbf{
\Large{Reactive Marketing and the Co-Production of (In)Authenticity}}
\vspace{0.2in}
\begin{table}[h!]
    \centering
    \begin{tabular}{c@{\hskip 0.8in}c}
         \textbf{Preyas S. Desai}   & \textbf{Jessie Liu}\\
         \textbf{Duke University}  & \textbf{Johns Hopkins University}\\
         (desai@duke.edu) &(jzliu@jhu.edu)\\
    \end{tabular}
    \end{table}

\vspace*{\stretch{1}}
{\centerline {\textbf{This Draft: November 2025}}} 
\vspace{0.2in}
{\centerline {\textbf{Abstract}}} 
\vspace{0.2in}
\noindent Businesses often react to external events by sending pro-social messages on social media that show the sender's alignment with the underlying prosocial cause and enhance their brand image. Consumers are uncertain about the authenticity of such messages because a company can choose to send prosocial messages even when their alignment with the social cause is not genuine.  We study the sender's incentives to send (in)authentic messages and the consumer's reactions when an external investigator can verify the sender's message.  
\bigskip{}

\noindent We find that the sender's equilibrium strategy depends on the receiver's emphasis on external investigation versus their self-signaling incentives. When the receiver is more internally focused, the sender chooses {\emph{self-sufficiency}} strategy, building credibility independent of external investigation.  When the receiver is more externally focused, the sender can either use self-sufficiency or {\emph{complementarity}}, the latter relying on validation by the external investigator. Thus, authenticity is coproduced by the sender, the receiver, and the investigator.  Importantly, self-sufficiency results in more authenticity in reactive marketing messages than complementarity.  

\bigskip{}

\noindent We extend the model to incorporate confirmation bias of the receiver.  We show that confirmation bias can act a doubled-edged sword, sometimes making the sender's persuasion task easier and other times making it more difficult. 
\vspace{\stretch{2}}

\noindent\rule{5in}{0.4pt}\\
\noindent \footnotesize{The authors are listed alphabetically. We acknowledge helpful comments from Kurt Carlson, Jack Soll, Chad Kendall, and participants in the Duke-UNC Seminar Series. The usual disclaimer applies.}
\end{titlepage}

\newpage
\setlength{\parindent}{0pt}
\section{Introduction}
\label{sec: intro}

Around the world, many consumers increasingly expect companies not only to provide the goods and services they seek, but also to take public stances on prosocial issues such as environmental sustainability, human rights, and social justice. The 2022 Edelman Trust Barometer survey \citep{edelman2022trustbarometer} found that nearly 60\% of global consumers make purchasing decisions based on brand values. 
%Similarly, the 2023 Edelman survey \citep{edelman2023trustbarometer} revealed that 63\% of respondents preferred brands that align with their own values. 
Similar results have been reported in other studies by Accenture \citep{accenture2018purposeled}, IBM \citep{ibm2022consumerstudy}, and McKinsey \citep{mckinsey2023sustainabilitywallet}.  In response to these expectations—and sometimes due to their own institutional values—companies and business leaders frequently issue public statements, including press releases and social media posts, when significant external events occurs. These statements may include messages about publicly recognized celebratory or awareness  occasions such as the Earth Day, Pride Month, World Mental Health Day, Asian American and Pacific Islander Heritage Month or unfortunate societal events. We refer to such messages as reactive messages.  

While consumers do not like inauthentic messages \citep{silver2021inauthenticity,berman2022prosocial}, corporate messages are costless and declarative in the sense that consumers cannot ``kick the tires."  Moreover, an average consumer might not have the resources to discern whether a company's prosocial message reflects the company's genuine values and practices or is merely tokenism. Therefore, a company can have incentives to make prosocial statements not only when they have a genuine alignment with a given cause or event but sometimes even when they do not have such an alignment. In this paper, we examine a business's incentives to issue public statements in response to an external event. We recognize that such corporate messages take place in a behaviorally rich environment that affect the consumers' reactions to reactive message and companies' messaging strategies.

An important aspect of the demand-side environment that we consider is the presence of external investigators who have the ability to evaluate the firm’s message. Unlike consumers, an investigator can draw on publicly available information about the firm’s history and practices to assess whether its statements align with past behaviors and messages. This possibility of verification enriches the strategic environment for consumers and companies. Companies now have to grapple with independent verification of its pro-social messages that might influence consumers beliefs and their reactions to the reactive messages.

Consumers' prosociality is a complex phenomenon.  Several papers on prosocial behaviors (\citealp[e.g.][]{wedekind2000cooperation, penner2005prosocial}) show that people like to not only help others and society in general but also like to support others who engage in such behaviors.  Furthermore, consumers' desire to demonstrate their own prosociality to themselves gives rise to self-signaling \citep{benabou2006incentives,dube2017self}.\footnote{People's desire to support prosocial messages \emph{regardless of message authenticity} (\citealp[e.g.][]{harbaugh2007neural, grant2008does, batson2011four}) are observationally equivalent to and can be grouped with self-signaling behavior.} We examine how consumers' self-signaling incentives affect reactive messaging strategies.  

While the consumers' self-signaling incentives represent their attitude towards a specific pro-social cause, their attitude towards the specific business is also important.  We incorporate this aspect of consumer behavior as well. Specifically, in a subsequent extension of the main model, we allow consumers' information processing to exhibit confirmation bias—i.e., the tendency to interpret information in a way that is consistent with pre-existing (prior) beliefs. Several papers such as \citet{klayman1987confirmation, hoch1986consumer, deighton1988can, klayman1995varieties, russo2002individual} establish the effects of confirmation bias in a variety of settings, including persuasion.  Given that consumers might have prior experiences with the firms or business leaders issuing reactive messages, confirmation bias could affect how they engage with these messages. 

Our focus on communication strategy, along with firms' private information about authenticity, leads us to apply the Bayesian Persuasion framework \citep{kamenica2011bayesian} to characterize the equilibrium mix of authentic and inauthentic messages.{\footnote{We use terms firm, business, and sender interchangeably.  Similarly, we use consumer and receiver interchangeably. }} In the model section, we discuss why this framework is better suited for our research questions than the traditional signaling model.

Taken together, these elements---independent investigation and consumers' attitudes towards the pro-social cause and the sender---collectively create a complex demand-side environment that enable us to go beyond mere authenticity concerns in our analysis of reactive communication strategies. This richness is grounded in behavioral theory and aligns with observed patterns on social media, where consumers respond not only to what firms say but also to how messages stand up to independent assessment and how they fit with consumers' own values and prior beliefs.

To provide additional evidence for these features in our specific context, we also conducted a short survey of social media participants ($n=201$) that asked questions about their desire to support prosocial corporate messages, and how it is affected by: (i) Their concerns for authenticity of such messages, and (ii) The impact of independent fact-checking of such messages.  The survey questions and its results are described in Appendix A.\footnote { The survey was pre-registered with the Open Science Foundation.  The survey and its results are also available at \url{https://osf.io/9ekyj/?view_only=9ed8eafd596c4f97ade6a984cc86bc8f)}.} The survey results support key aspects of our model highlighted above.

\subsection{Overview of Model and Results}

We begin our analysis with a model in which a sender chooses the frequency with which they send prosocial messages to a receiver.  The frequency of messaging is separately defined for cases in which there is a good fit between an exogenous event and the sender's business and those in which the fit is not good. We refer to messages in the former case as authentic messages and those in the latter case as inauthentic messages.  We also model an independent investigator who can verify the authenticity of a given message.  The receiver's reaction, in terms of supporting the message or ignoring it, depends on the receiver's beliefs about the authenticity of the message, which itself is affected by the investigator's analysis and the receiver's self-signaling incentives.  We find that there is a set of parameters for which the receiver always supports the sender's message, regardless of the sender's strategy.  We refer to this set of parameters as the zone of \textit{automatic affirmation}. This zone of automatic affirmation expands not only with the priors about the sender, but also with the receiver's self-signaling incentives.  Outside the automatic affirmation zone, the sender needs to ensure that their message is sufficiently credible to the receiver.  Given the presence of the external investigator, the sender has two strategic options.  The first is to gain \textit{self-sufficiency} by reducing the frequency of inauthentic messages such that the receiver would find any given message credible even if the investigator negates it.  The second is a \textit{complementarity} strategy in which the sender leverages the external investigation to win the receiver's support. While a mix of authentic and inauthentic messages is observed under each strategy, the self-sufficiency strategy results in fewer inauthentic messages. This is paradoxical because in the self-sufficiency equilibrium, the external investigator has no material impact on the receiver's decision to support the sender's message but it is in this equilibrium that the sender is more authentic.  

In the spirit of thought experiments, we next consider the possibility of the receiver's confirmation bias.  We again find that both self-sufficiency and complementarity strategies can be observed in equilibrium.  The receiver's confirmation bias, however, can either facilitate or hinder the sender's task of persuading the receiver.  This is despite the fact that the receiver with confirmation bias gives less weight to the investigator.  Confirmation bias can be a double-edge sword for the sender and the receiver.

\subsection{Relationship with the Literature}

A large literature in economics, marketing, and management science has examined the spread of misinformation and disinformation, focusing on how false or misleading content propagates through networks and how individuals, knowingly or unknowingly, contribute to its diffusion \citep{candogan2017optimal, acemoglu2024model}. Another stream of work focuses on the production of misinformation \citep{wang2023news,godes2023will, ederer2022bayesian, yang2024strategic}. 
Like these studies, our paper focuses on strategic message generation rather than passive propagation. However, we depart from prior work by underscoring a co-production mechanism in which firms and consumers jointly sustain (in)authentic communication, even in the absence of ideological demand, and across varying levels of investigative accuracy. 

Fact-checking in \cite{godes2023will} functions as a partisan signal, motivating the production of media bias that caters to consumers’ ideological preferences. In contrast, investigation in our model enables both authentic and inauthentic communication—whereas the former is absent in \cite{godes2023will}. Furthermore, unlike \cite{ederer2022bayesian} and \cite{yang2024strategic}, who primarily focus on the optimal design of lie detection technologies, we emphasize how receivers’ desire to align their action with self-image and prior beliefs, both of which are central to the context of reactive marketing, affect their interpretation of the investigation outcome and, in turn, influence firms’ strategic behavior.

As we discussed earlier, our paper builds on a few well-known findings from psychology and economics. The literature on self-signaling \citep{quattrone1984causal, bodner2003self, benabou2006incentives} shows that sometimes people strategically take actions that later serve as signals to themselves about their own values and beliefs.  Papers such as \cite{gneezy2012pay} and \cite{dube2017self} have also shown self-signaling reasons for prosocial behaviors. Behavioral research complements this by showing that consumers are sensitive to perceived inauthenticity, especially in prosocial or moral domains  \citep{silver2021inauthenticity, karikari2024three}.  Moreover, several papers on confirmation bias (see \cite{klayman1995varieties} for a review) and pre-decisional distortion (see \cite{russo2014predecisional} for a review) show that people often overweight their existing beliefs over new information. These studies typically explore how individuals' attitudes and beliefs  shape their own decisions. We use self-signaling and the preference for authenticity as building blocks for our model and then examine how they impact the effectiveness of external validation and the sender’s strategic behavior. 

Third, our work builds on the Bayesian persuasion literature, which analyzes how information design can influence rational decision-makers \citep{kamenica2011bayesian}. 
Recent applications to marketing have explored persuasion in contexts such as product affiliation, advertising, news provision, and notification strategies \citep[e.g.,][]{pei2022influencing, iyer2022pushing, wang2023news, yao2024dynamic, shin2024role}. While these models generally assume Bayesian receivers, recent work has begun to examine the implications of non-Bayesian belief updating on the effectiveness of persuasion \citep{de2022non}. Our paper contributes to this literature by focusing on corporate communication, where the sender’s (firm's) authenticity is uncertain and the receivers' psychological motivations are endogenous. An instrumental feature of our model is the presence of an external investigator, which introduces a second channel of belief updating that reshapes the co-production incentives of both the sender and receiver. This leads to two distinct communication strategies—complementarity and self-sufficiency—through which firms manage credibility in response to potential validation or contradiction. These strategies emerge under Bayesian updating, but we also show how confirmation bias further shapes these strategies in a non-Bayesian setting.

Finally, our paper contributes to the growing literature on corporate stance-taking, which investigates how consumers respond to companies that engage with social or political issues. This literature has primarily focused on measuring consumer reactions, using observational data or experiments to estimate changes in sales \citep{liaukonyte2023goya,painter2021walmart,hydock2020firmstances, conway2024consuming}, brand favorability \citep{klostermann2021brandindex}, or digital engagement \citep{schoenmueller2023twitter}. 
While these papers shed light on the consequences of corporate messages, they typically treat firms' communication choices as exogenous. In contrast, to the best of our knowledge, our paper is the first to model stance-taking as an \textit{endogenous} outcome: firms anticipate consumer preferences, biases, and the presence of external validation when deciding whether and how often to take a stance. Moreover, whereas previous research has focused on stance-taking around controversial issues, our emphasis is less on polarization per se than on the firm's credibility challenge: the pressure to react to broadly appealing social causes, and the temptation to \textit{appear} authentic without genuine alignment. Even when an issue is not polarizing, firms need to deal with authenticity questions.  In this respect, our paper is also related to \citet{wu2020bad}, where authenticity can be signaled through costly observable (vs. non-observable) investments. In our setting, however, all investments are non-observable to consumers and communication takes the form of ex ante costless messages. Although consumers cannot be completely certain, they might be persuaded by a message under some conditions.  The persuasiveness of a message emerges from the strategic interaction between the firm, the investigator, and the consumer. Firms expect that consumers seek self-signaling value from supporting the cause but rely on external validation to assess authenticity. This expectation shifts the firm’s problem from signaling virtue through transparent, costly action to sustaining persuasion through declarative messages, as is often the case in corporate stance-taking.

The remainder of the paper is organized as follows. Section \ref{sec: model} introduces the model. Section \ref{sec: analysis} analyzes the baseline case without confirmation bias, while Section \ref{sec: CB} examines how confirmation bias shapes equilibrium outcomes. 
Section \ref{sec: conclusion} concludes. 

\section{Model} \label{sec: model}

In this section, we describe our model and its key assumptions. Before presenting our formal model, we outline the context for our model in terms of reactive messages and the sender's strategic problem. 

\subsection{Strategic Context and Modeling Framework} We consider a sender choosing its reactive marketing strategy in terms of whether and when to send prosocial messages through a public communication channel. The sender could be a business or a leader of the organization.  The receiver of the communication is a consumer, or more generally, a stakeholder, who decides whether or not to support the sender upon seeing the message.\footnote{Throughout the paper, we use they/them pronouns for all agents.} 

While senders may engage in public messaging on a variety of topics, including about their products, our focus is \emph{reactive} messages that are issued in response to exogenous events. Such events include social problems, natural disasters, or public commemorations, such as a day or week honoring a specific affinity group. Reactive messages serve to express the sender’s alignment with or advocacy for the prosocial cause connected with the exogenous event. By issuing such messages, the sender can position themselves as a responsible member of the society. It also helps them meet the customers’ expectations that they take stand on social issues. 

Importantly, we do not view a sender’s \emph{true} alignment with a cause as a reflection of their moral character. Instead, we think of it as a consequence of how strongly the cause is connected to the sender’s business and/or the sender's past actions to support it.

Although a sender can proclaim affinity for a cause through a reactive message, the message itself is intangible and not conducive to ``kicking the tires." Further, an average receiver does not have the resources to verify the authenticity of the message. This creates the central strategic question for the sender: Should they restrict themselves to sending reactive messages only when they have a genuine commitment to the underlying cause or should they also espouse causes for which they themselves do not have a genuine commitment.

With this background, it might seem that both a traditional signaling model and the Bayesian Persuasion model could possibly be used to analyze the sender's strategy.  However, as we explain below, the Bayesian Persuasion Model is better suited for our purposes.  

First, the sender's main strategic choice is about corporate \emph{messages}, which is essentially \emph{an information design problem}.  Second, the action that the sender takes, viz., sending out a message is in itself not a costly action.  This also makes the traditional signaling model a less preferred option.  Third, the receiver in our context is an ordinary consumer.  Although we recognize that an analytical model cannot be descriptively correct in all respects \citep{friedman1953essays}, the informational and epistemic assumptions required for signaling models to work are more strenuous for our context than those required for the Bayesian Persuasion model.  For example, the receiver in the signaling model would need to correctly invert every sender type's strategies and payoffs, correctly deal with off-equilibrium beliefs, and apply complex refinements such as the Intuitive Criterion. 

While cheap talk does model costless messages, it is not suited for our analysis because it assumes that these messages are non-verifiable, which might only lead to uninformative babbling equilibria. In reality, corporate statements often do shift posterior beliefs precisely because they are cross-checked by third parties, making complete babbling implausible. Moreover, cheap talk models characterize equilibrium message choices \emph{ex post}, given beliefs, but the central trade-off in our setting involves the \emph{ex-ante design} of an information structure.  

In Section 6, we discuss the familiar question about message commitment associated with Bayesian Persuasion models and a potential resolution.

We now describe the sender's strategic problem more formally.

\subsection{Sender and Reactive Messages} 
We adopt the Bayesian Persuasion framework \citep{kamenica2011bayesian} for our analysis. The sender’s cumulative investments effectively determine their ``fit'' with a given cause in our model. For a given cause, the sender's fit can take one of two values, denoted by $\theta \in \{0,1\}$. When the sender has a good (strong) fit, $\theta = 1$ and when the sender does not have a good fit, $\theta = 0$.  The parameter $\theta$ captures the quality of fit between the sender and the cause, rather than making any normative judgment about the sender. Let $\rho_0\in[0,1]$ denote the prior probability that $\theta = 1$.

The sender chooses a pair $\{r_G, r_B\}$, ($r_G \in [0,1], r_B \in [0,1]$), which are the probabilities of sending out a prosocial message when they discover their type to be $\theta=1$ and $\theta=0$, respectively, following an exogenous event. We define the sender's message, $m\in\{0,1\}$ such that $m=1$ refers to a prosocial reactive message being sent by the sender and $m=0$ refers to no such message being sent. Thus, an important aspect of our model is that the firm may choose not to send any reactive marketing message. In our model, the probability of a reactive message being sent is $\rho_0 r_G + (1-\rho_0) r_B$.

The sender’s message, $m$ is considered ``authentic'' when $m(\theta=1)=1$ and  ``inauthentic'' when $m(\theta=0)=1$. Authenticity is not defined for $m=0$ because no message is sent in this case. Accordingly, the proportions of authentic vs. inauthentic messages among all messages being sent are given by $\frac{\rho_0r_G}{\rho_0r_G+ (1-\rho_0)r_B}$ and $\frac{(1-\rho_0)r_B}{\rho_0 r_G+ (1-\rho_0)r_B}$, respectively.

As we noted above, the sender's goal in sending the reactive prosocial message is to obtain the receiver's support and meet the latter's expectations of the sender.  
Denoting the sender's objective function by $\pi^s$, we assume that $$\pi^s =\mathbf{1}(a) \cdot \underline{\pi},$$ where $\underline{\pi}>0$ is the unspecified utility of getting a positive brand image, and $\mathbf{1}(a)=1$ if $a=1$, i.e., the receiver supports the sender, and $0$ otherwise. We normalize $\underline{\pi}$ to 1 for simplicity. In other words, $\pi^s$ is the incremental profit from sending a prosocial message in reaction to an exogenous event and not from the entirety of the sender's social media activities.  

\subsection{Receiver} 

As the above discussion implies, the cause that we consider is one for which there is no disagreement among the sender and the receiver.  Further, we allow for the possibility that a socially-conscious receiver gets a positive value from endorsing a prosocial cause that they believe in.  This potential positive value can come from the receiver's altruism or pro-social attitudes \citep{small2016prosocial, white2020review} or from the receiver's self-signaling needs \citep{benabou2006incentives}.
%\footnote{This is how it fits the self-signaling framework.}  
On the other hand, the receiver also recognizes that the sender might only be engaging in tokenism or window dressing in sending the message, i.e., $\theta=0$.  The receiver prefers to avoid supporting the message in such cases \citep{small2016prosocial}.  We incorporate these two elements into the receiver’s utility function, $U$, defined as:
\[U =
\begin{cases}
u(a) - (a - \theta)^2, & \text{if } m = 1 \\
0, & \text{if } m = 0
\end{cases},\]
where $a \in A(m)$ denotes the receiver’s binary decision to support the message ($a = 1$) or not ($a = 0$). Support may be manifested through engagement with the sender's (business's) social media content — for example, through likes, shares, and comments. The receiver is only active when a message is sent; that is, the action set is defined as:
\[A(m) =
\begin{cases}
\{0, 1\}, & \text{if } m = 1 \\
\emptyset, & \text{if } m = 0
\end{cases}.\]
The receiver's action space is limited to supporting or not supporting a reactive marketing message.  Therefore, when there is no message from the sender, the question of the receiver supporting an action does not arise and the receiver does not take any action.  We recognize that the receiver may take other actions to support the sender, but our interest is in studying the receiver's support on social media.\footnote{Additional actions by the receiver can be modeled by including the benefits of such actions to the sender and the receiver and the cost of such actions to the sender.}  

We assume that $u(a=1)>u(a=0)$ and define the self-signaling incentive as $v \equiv u(a = 1) - u(a = 0) \in [0,1]$. When $v = 0$, the receiver derives no utility from supporting the message itself. Finally, we assume a tie-breaking rule in favor of the sender in that when indifferent between supporting and not, the receiver chooses to support the sender's message (i.e. $a=1$).  We consider only one receiver because we want to focus on the receiver's support decision in isolation of social signaling or popularity effects.{\footnote{In Appendix, we consider a case that accommodates multiple receivers but without social signaling.}}

\subsection{Investigator}
We also incorporate an investigator as another participant on the social media. The investigator in our model is present on the same social media platform and likes to verify the social media messages by senders. Unlike the receiver, the investigator has the time to do additional research about the sender and evaluate if the sender's message is authentic or inauthentic.  They can do this by reading previous messages by the sender and looking up publicly available information about the sender that can reveal whether or not the sender's (investment) actions and previous messages reveal a sincere commitment to the underlying cause. To simplify our analysis and focus on the receiver's and the sender's decisions, we assume that the investigator derives intrinsic utility from conducting and publicizing an investigation that exceeds their cost of doing so. Moreover, the investigation occurs only if a prosocial message ($m=1$) is sent, since the signal is designed to verify the authenticity of that message. Therefore, no investigation takes place if the sender has not sent a message.  

The outcome of the investigation is denoted by $s=\{0,1\}$, where $s=1$ indicates evidence suggesting the sender is a good type, and $s = 0$ indicates evidence suggesting the sender is a bad type. In other words, $s = 1$ signals the presence of evidence for genuine investment, while $s = 0$ reflects a lack of such evidence.{\footnote{In practice, investigators could also resemble features like an ``AI fact-check'' button, automated content moderation tools, or community-driven systems such as Twitter's (now ``X'') Community Notes, which flag or contextualize questionable claims.}}

We recognize that the investigator's research is not perfect and could have (non-systematic) errors.  We denote the probability of the investigator's research (and thus the subsequent signal) correctly assessing a good type sender as a good type by $p$.  That is, $\mathcal{P}(s=1|\theta=1)=p$. The probability of the investigator's signal incorrectly showing a bad type as a good type by $q$, so that $\mathcal{P}(s=1|\theta=0)=q$. The parameter $p$ captures the probability that an investigator finds true evidence when the firm has genuinely invested; $q$ reflects the probability that (misleading) evidence appears even when no investment occurred. We assume that $0<q<\frac{1}{2}$ and $\frac{1}{2}<p<1$ so the investigation is informative.\footnote{$p=q=\frac{1}{2}$ corresponds to the case of uninformative investigation.}  These values of $p$ and $q$ are known to the investigator, sender, and receiver. Because our focus is on strategic misrepresentation by the sender, we assume that the investigator truthfully reports the outcome of their investigation.  In other words, the investigator does not create strategic distortion in their signal.  This could be due to the investigator's concern for factual information and/or their reputation as investigator. 

In the absence of any messaging or investigation, the receiver and the investigator believe that the sender is of a good type ($\theta=1$) with probability $\rho_0\in[0,1]$ and of a bad type ($\theta=0$) with probability $1-\rho_0$.  We discuss the belief updating process in the next subsection.

\subsection{Game Structure and Bayesian Updating}
  The game proceeds in four stages. In the first stage, the sender has not discovered their true type and chooses their overall communication strategy for the entire planning horizon. Specifically, the sender chooses a pair $\{r_G, r_B\}$, ($r_G \in [0,1], r_B \in [0,1]$), which are the probabilities of sending out $m=1$ message when they discover their type to be $\theta=1$ and $\theta=0$, respectively. In Stage 2, an exogenous event occurs, the sender discovers their true type and implements their committed messaging strategy. In Stage 3, the receiver and the investigator observe the message. The investigator conducts their investigation and sends a signal $s=\{0,1\}$.\footnote{In the remainder of the paper, we refer to the investigator's message as the signal and the sender's message as the message.} In Stage 4, the receiver observes the investigator's signal and chooses whether or not to support any prosocial message from the sender.

Figure 1 describes the sequence of events.

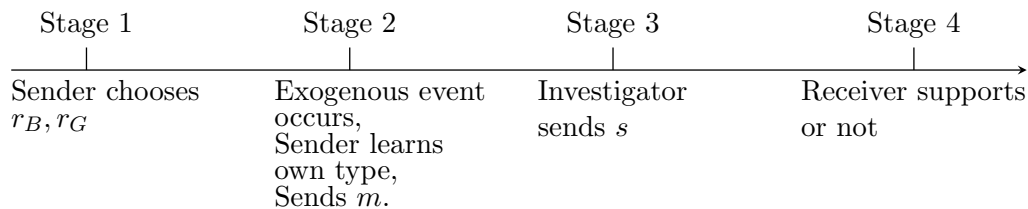
\begin{figure}[h!]
	\centering
	\begin{tikzpicture}
		\usetikzlibrary{calc}
		% draw arrow
		\coordinate (start) at (-11.5,0);
		\coordinate (end) at (2,0);
		\draw [line width=0.5pt, -stealth] (start) -- (end);
		
		% You can use `foreach` to improve the following codes
		\coordinate (s0) at (-10.5,0);
		\coordinate (w0) at (-10,0);
		\coordinate (t0) at ($(s0)+(0,0.3)$);
		\coordinate (s1) at (-7,0);
		\coordinate (w1) at (-6.5,0);
		\coordinate (t1) at ($(s1)+(0,0.3)$);
		\coordinate (s2) at (-3.5,0);
		\coordinate (w2) at (-3,0);
		\coordinate (t2) at ($(s2)+(0,0.3)$);
		\coordinate (s3) at (0.5,0);
		\coordinate (w3) at (0.0,0);
		\coordinate (t3) at ($(s3)+(0,0.3)$);

		% draw ticks
		\draw [line width=0.5pt] (s0) -- (t0);
		\node [anchor=south] at (t0.north) {Stage 1};
		
		\draw [line width=0.5pt] (s1) -- (t1);
		\node [anchor=south] at (t1.north) {Stage 2};
		
		\draw [line width=0.5pt] (t2) -- (s2);
		\node [anchor=south] at (t2.north) {Stage 3};
		
		\draw [line width=0.5pt] (t3) -- (s3);
		\node [anchor=south] at (t3.north) {Stage 4};
		
		% add texts
		\node [anchor=north, align=left, text width= 3cm] at (w0.south) {\baselineskip=10pt
		Sender chooses $r_B, r_G$

		};
		
		\node [anchor=north, align=left, text width=3cm] at (w1.south) {\baselineskip=10pt
			Exogenous event occurs, \\ Sender learns own type, \\ Sends $m$.
			
		};
		
		\node [anchor=north, align=left, text width=3cm] at (w2.south) {\baselineskip=10pt
						Investigator sends $s$
			};
		
		\node [anchor=north, align=left, text width=3cm] at (s3.south) {\baselineskip=10pt
			Receiver supports or not
					};
		
	\end{tikzpicture}
	\caption{Sequence of Events}\label{fig: timing}
\end{figure}

Consistent with the Bayesian Persuasion framework, we assume that the receiver updates their prior beliefs, $\rho_0$, upon observing the sender's message, $m$, in Stage 2.  This updated belief, $\rho_1$, is given by 
\begin{equation}
 \rho_1(m=1) = 
\frac{r_G \rho_0}{r_G \rho_0 + r_B (1-\rho_0)} \label{eq: rho1}
\end{equation}
As the above equation shows, the receiver takes into account the sender's optimal strategy, $\{r_G, r_B \}$, in forming $\rho_1$.

The receiver's updating of beliefs does not end here as they also consider the investigator's signal when it becomes available in Stage 3.  

The receiver's posterior beliefs, $\rho_2$, after observing the signal (as well as the message) are given as follows:
\begin{equation}
\rho_2(s) = 
\begin{cases} 
\frac{p \rho_1}{p \rho_1 + q (1 - \rho_1)} & \text{if } s = 1 \\[10pt]
\frac{(1 - p) \rho_1}{(1 - p) \rho_1 + (1 - q)(1 - \rho_1)} & \text{if } s = 0
\end{cases}
\label{eq: rho2}
\end{equation}
Note that $\rho_1$ is a function of message $m$ which arrives before the signal.

Next, we describe our analysis and key results. Proofs for all results are provided in the Appendix.

\section{Analysis} \label{sec: analysis}

\subsection{Receiver's Strategy}

In Stage 4, the receiver will support the sender's prosocial claim if and only if their posterior belief about the sender being a good type exceeds a specific threshold\footnote{Note that 
\begin{align*}
U\left(a=1\right)-U\left(a=0\right)	& =v-\mathbb{E}\left[\left(1-\theta\right)^{2}\right]-\left(0-\mathbb{E}\left[\left(0-\theta\right)^{2}\right]\right) \\
	& =v-\mathbb{E}\left[\left(1-\theta\right)^{2}-\theta^{2}\right] \\
	& =v-\mathbb{E}\left[1-2\theta\right].
\end{align*}
Therefore, we have $U\left(a=1\right)-U\left(a=0\right)\geq0\Longleftrightarrow\mathbb{E}\left(\theta\right)=\rho_{2}\geq\frac{1}{2}\left(1-v\right)$.}, i.e., 
\begin{equation}
 \mathbf{\rho_{2}}\left(m,s\right)\geq \frac{1}{2}(1-v) \label{eq:rho2cutoff}  
\end{equation}
 where $v$ captures the utility of self-signaling.

Combining equations \eqref{eq: rho1} and \eqref{eq: rho2}, we can write the receiver's posterior belief, $\mathbf{\rho_{2}}\left(m,s\right)$, in stage 4 as follows: 
\begin{align}
\mathbf{\rho_{2}}\left(m=1,s=1\right)	&=\frac{pr_{G}\rho_{0}}{q\left(1-\rho_{0}\right)r_{B}+pr_{G}\rho_{0}} \label{eq:rho2m1s1}\\
\mathbf{\rho_{2}}\left(m=1,s=0\right)	& =\frac{\left(1-p\right)r_{G}\rho_{0}}{\left(1-q\right)\left(1-\rho_{0}\right)r_{B}+\left(1-p\right)r_{G}\rho_{0}} \label{eq:rho2m1s0} %\\
%\mathbf{\rho_{2}}\left(m=0,s=1\right)	& =\frac{p\left(1-r_{G}\right)\rho_{0}}{q\left(1-\rho_{0}\right)\left(1-r_{B}\right)+p\left(1-r_{G}\right)\rho_{0}} \\
%\mathbf{\rho_{2}}\left(m=0,s=0\right)	& =\frac{\left(1-p\right)\left(1-r_{G}\right)\rho_{0}}{\left(1-q\right)\left(1-\rho_{0}\right)\left(1-r_{B}\right)+\left(1-p\right)\left(1-r_{G}\right)\rho_{0}}.
\end{align}

Before going forward, we note that Equations \eqref{eq:rho2cutoff}-\eqref{eq:rho2m1s0} show the joint impact of the receiver's self-signaling incentives and the results of the external investigation on the receiver's decision to support or not support the sender's message. 

Let $\rho(s)$ denote the receiver's posterior belief about the sender's type $\theta$ based solely on the signal $s$, conditional on the message $m=1$ being completely uninformative, i.e., $\rho_1=\rho_0$. In particular, we have $\rho\left(s=1\right)=\frac{\rho_{0}p}{\rho_{0}p+\left(1-\rho_{0}\right)q}$ and $\rho\left(s=0\right)=\frac{\rho_{0}\left(1-p\right)}{\rho_{0}\left(1-p\right)+\left(1-\rho_{0}\right)\left(1-q\right)}$. This leads to the following lemma.  

\medskip{}

\begin{lemma}\label{lem: posterior}
$\mathbf{\rho_{2}}\left(1,s\right)\geq\rho\left(s\right),\forall s\in\left\{ 0,1\right\}$
with the equality holding only when $m$ is not informative at all, i.e., when $r_{G}=r_{B}$.
\end{lemma}

First, note that the prosocial message is sent weakly more often when the sender is a good type than a bad type ($r_G^*\geq r_B^*$) in any equilibrium. This is because, in order to increase the likelihood of receiver's support, the sender needs to maximize the occurrence of reactive messages meanwhile maintaining their credibility.  Then Lemma \ref{lem: posterior} follows immediately from Equations \eqref{eq:rho2m1s1} and \eqref{eq:rho2m1s0} given that $r_G^*\geq r_B^*$.\footnote{See Appendix B for a formal proof of Lemma \ref{lem: posterior}.}  

\subsection{Sender's Strategy}
The sender's strategy in our model consists of choosing probabilities for sending reactive marketing messages under the two states.  We find that the sender's equilibrium choices depend on the receiver's self-signaling incentives relative to the prior beliefs and the investigative technology.   
We begin with the characterization of conditions under which the veracity of the sender’s message does not affect the receiver's decision to support the sender. We refer to this case as automatic affirmation by the receiver.     

\begin{proposition}
\textbf{(Automatic Affirmation)}
\label{prop: auto}
There exists a level of prior belief,  $\overline{\rho}\equiv\frac{\left(1-v\right)\left(1-q\right)}{\left(1-v\right)\left(1-q\right)+\left(1+v\right)\left(1-p\right)}< 1$, such that for $\rho_0>\overline{\rho}$  the sender always sends the prosocial message regardless of their type, i.e., $r_G^*=r_B^*=1$, and the receiver always supports them. Moreover, $\frac{\partial \overline{\rho}}{\partial v}<0$ and $\frac{\partial \overline{\rho}}{\partial p}>0$, but $\frac{\partial^2\overline{\rho}}{\partial v \partial p}<0$ if $v>\frac{p-q}{2-p-q}>0$ and $\frac{\partial^2\overline{\rho}}{\partial v \partial p}\geq0$ if $0<v\leq \frac{p-q}{2-p-q}$.
\end{proposition}

When either the receiver's prior beliefs or their self-signaling incentives are sufficiently high, the receiver is predisposed to support a prosocial message from the sender.  When $\rho_0>\overline{\rho}$, the receiver supports the prosocial message from the sender, even when they recognize that the sender's message could be inauthentic.  Furthermore, the receiver supports the sender's message even when the investigator's signal is negative. In such cases, the sender does not need to be concerned about their credibility and always delivers a prosocial message to further their brand-building or reputational goals. Although we describe Proposition 1 in terms of $\rho_0$, the parameter space for automatic affirmation by the receiver is defined not only by $\rho_0$ but also by the receiver's self-signaling utility as well as properties of the investigator's signal. The cutoff value of this prior belief, $\overline{\rho}$, depends on the accuracy of the investigator's signal, increasing either with higher $p$ or lower $q$. The cutoff value also decreases with a stronger self-signaling incentives ($v$), and this decline is more pronounced when $p$ is high. Ironically, as the investigator's signal becomes more accurate, it exerts \textit{less} influence on the decision of receivers — not due to a lack of informational value, but precisely because of it: the signal is strategically invoked to validate a choice receivers were already predisposed to make when their self-signaling incentives are strong.

We now turn to cases in which the condition in Proposition \ref{prop: auto} is not satisfied and the receiver's response and the sender's strategy are determined by the informativeness of the sender's message and the investigator's signal.  As before, the sender needs to send a message to have \emph{any} chance of receiving the receiver's support.  But in the absence of automatic affirmation, the sender must choose $\{r_G^*, r_B^*\}$ such that the receiver finds the message credible with a sufficiently high probability, i.e., $\rho_2(m, \cdot) \geq \frac{1-v}{2}$.  

We find that $r_G^*=1$ even when $\rho_0<\overline{\rho}$  because the sender has no incentives to be silent in the good state. However, the sender's choice of $r_B^*$ is not directly apparent. 
In particular, the receiver's self-signaling incentives as well as the possibility that an external investigator can validate or dispute the sender's message both add additional strategic considerations for the sender. These two factors lead the sender to adopt one of the following two strategies: (i) A \textit{complementarity} strategy, where the sender relies on corroboration by the investigator, and (ii) A \textit{self-sufficiency} strategy, where sender makes their message persuasive on its own. In the first case, the sender's message and the investigator's signal are complementary  whereas in the second case, the sender's message is sufficient to persuade the receiver and the investigator's signal ends up being inconsequential.

It turns out that the sender's choice depends on the quality of the investigative technology relative to the receiver's self-signaling incentives, specifically on whether $p \leq \overline{p}=\frac{2-(1-v)q}{3-2q+v}$ or $p>\overline{p}$. The receiver relies less on the investigator's validation in the former case than in the latter case. Therefore, we interpret these two cases as reflecting different emphasis in the receiver's decision-making process and refer to them as internal focus and external focus, respectively. Please note that these labels are meant to convey a \emph{relative emphasis} between self-signaling and external investigation, and not a dichotomy.     

\begin{proposition} \label{prop: comp-and-self} 
When $\rho_0<\overline{\rho}=\frac{\left(1-v\right)\left(1-q\right)}{\left(1-v\right)\left(1-q\right)+\left(1+v\right)\left(1-p\right)}$, 
\begin{enumerate}[leftmargin = 1.5em, label=(\roman*)]
\item \textbf{Receiver Internal Focus} ($p\leq \overline{p}$): The sender adopts the self-sufficiency strategy,  given by 
$r_B^*=r_B^{\text{self}}$.
\item \textbf{Receiver External Focus} ($p>\overline{p}$): The sender adopts the complementarity strategy, $r_B^*=r_B^{\text{comp}}$, for $\rho_0<\hat{\rho}$, and then switches to the self-sufficiency strategy, $r_B^*=r_B^{\text{self}}$, for $\hat{\rho}\leq \rho_0<\overline{\rho}$, 
\end{enumerate}
where $r_B^{self}=\left(\frac{1-p}{1-q}\right)\left( \frac{1+v}{1-v}\right)\left(\frac{\rho_0}{1-\rho_0} \right)$,   $r_B^{comp}=\min\left\{\left(\frac{p}{q}\right)\left( \frac{1+v}{1-v}\right)\left(\frac{\rho_0}{1-\rho_0} \right), \; 1\right\}$,\\ and $\hat{\rho}=\frac{(1-q)q(1-v)}{\left(p-q\right)q(1-v)+2\left(1-p\right)}$. 
\end{proposition}

When the receiver's focus is internal, ($p \leq \bar{p}$), the receiver considers the investigator's signal but places a lower weight on it (see Equations \eqref{eq:rho2cutoff}-\eqref{eq:rho2m1s0}). From the sender’s perspective, this limits the usefulness of positive validation by the investigator, leading them to adopt a self-sufficiency strategy. They choose $r_B^* > 0$ but keep it low enough to build their credibility independently. The sender's goal here is to persuade the receiver to offer support even when the investigator’s signal contradicts the sender’s message. While one might expect that the external threat of contradiction would deter inauthentic messages, the opposite happens: the sender is \textit{more} disciplined when the external investigation is less diagnostic for the receiver.

In contrast, the case of receiver's external focus  ($p > \bar{p}$) creates strategic opportunities for the sender. As the sensitivity of signal becomes higher, the value of receiving a validating signal ($s = 1$) increases, thereby enhancing the credibility of the sender's message.\footnote{When $s = 1$, the sender's message is validated by the investigator, and the receiver's posterior belief is given by: $\rho_2(s = 1) = \frac{p \rho_1}{p \rho_1 + q (1 - \rho_1)}$. This posterior belief increases with $p$.} When the prior beliefs are low, the sender lacks intrinsic credibility and thus adopts a complementarity strategy, choosing $r_B^* = r_B^{\text{comp}}$ and relying on the investigator's validation to convince the receiver. 

As $\rho_0$ increases, the sender becomes more credible and increases $r_B^*$. However, once $\rho_0$ surpasses a critical threshold $\hat{\rho}$, the sender reverts to the self-sufficiency strategy ($r_B^* = r_B^{\text{self}}$), as the positive value of the external validation is reduced.  At this stage, it is better for the sender to reduce $r_B$. 

Proposition 2 further shows that the relative magnitudes of the internal versus external factors ($v$ relative to $p$ and $q$) also determine the equilibrium values of $r_B^*$. Moreover, it can be seen that $r_B^{\text{self}} < r_B^{\text{comp}}$, holding all else constant.

Taken together, these results show that authenticity and inauthenticity in reactive messaging are not merely fixed traits nor are they purely sender-driven outcomes. Rather, they are strategic outcomes shaped through the interplay of three forces: the sender’s incentives to craft a persuasive message, the credibility of investigation signals, and the receiver's desire to interpret information in ways that align with their self-image. Thus, authenticity and inauthenticity, are co-produced, emerging from the equilibrium actions of all parties involved.

\begin{figure}[h!]
    \centering
    \begin{subfigure}[t]{0.48\textwidth}
        \centering
        \includegraphics[scale=0.4,trim=1cm 6.5cm 1.2cm 6.5cm,clip]{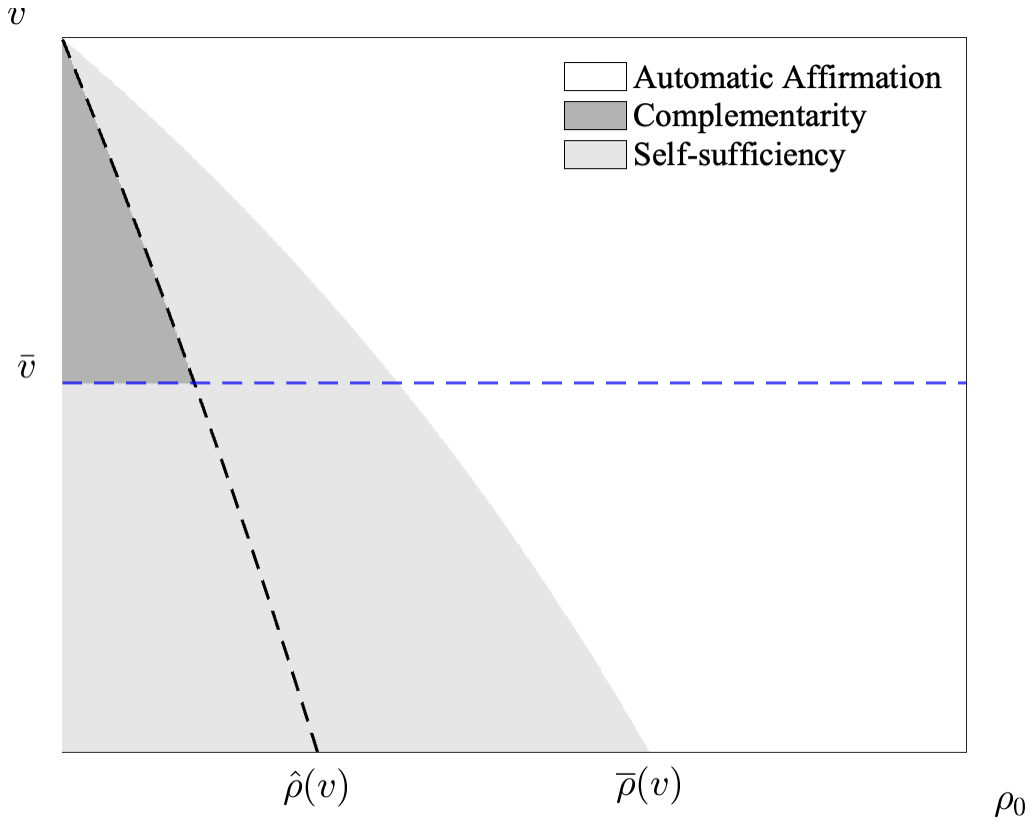}
        \caption{$p\leq \overline{p}(v=0)=\frac{2-q}{3-2q}, (p=0.65,q=0.35)$;}
        \label{fig: low-accuracy}
    \end{subfigure}
    \hfill
    \begin{subfigure}[t]{0.48\textwidth}
        \centering
        \includegraphics[scale=0.4,trim=1cm 6.5cm 1.2cm 6.5cm,clip]{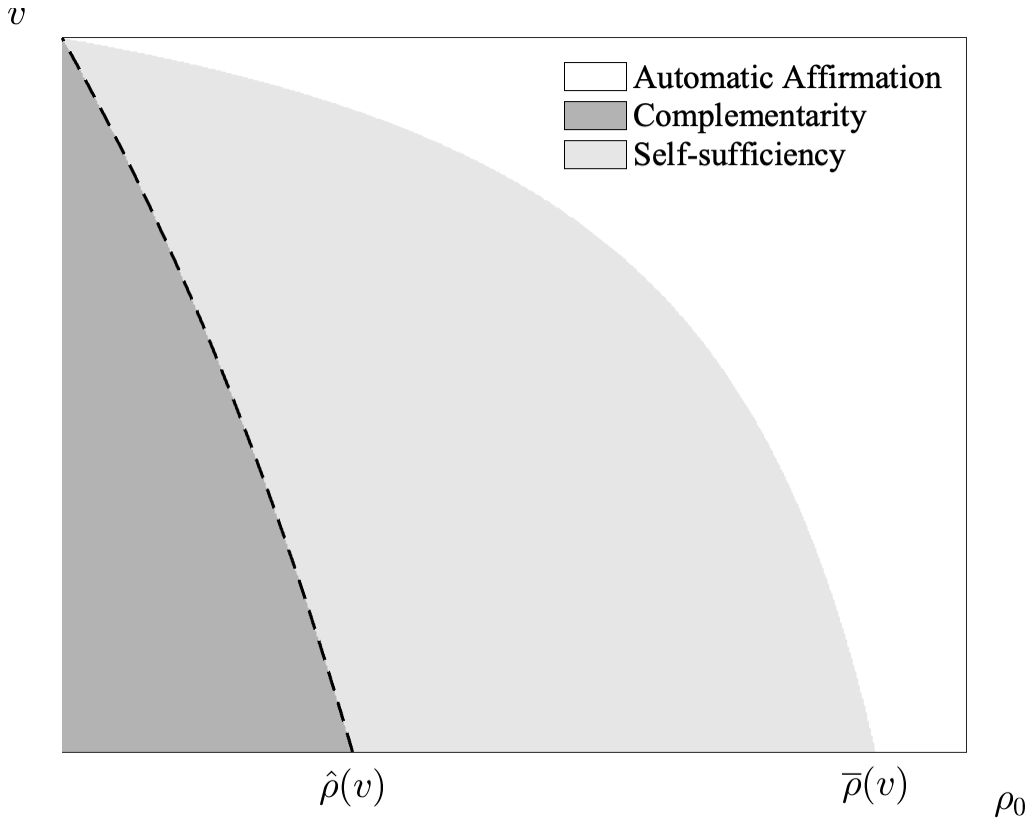}
        \caption{$p>\overline{p}(v=0)$, $(p=0.9,q=0.1)$}
        \label{fig: high-accuracy}
    \end{subfigure}%
    \caption{Coproduction of Authenticity: Equilibrium Strategies in the $v$-$\rho_0$ Space} 
   
    \label{fig:eq-region-rho-vs-v}
\end{figure}

\begin{figure}[h!]
    \centering
    \begin{subfigure}[t]{0.48\textwidth}
        \centering
        \includegraphics[scale=0.4,trim=1cm 6.5cm 1.2cm 6.5cm,clip]{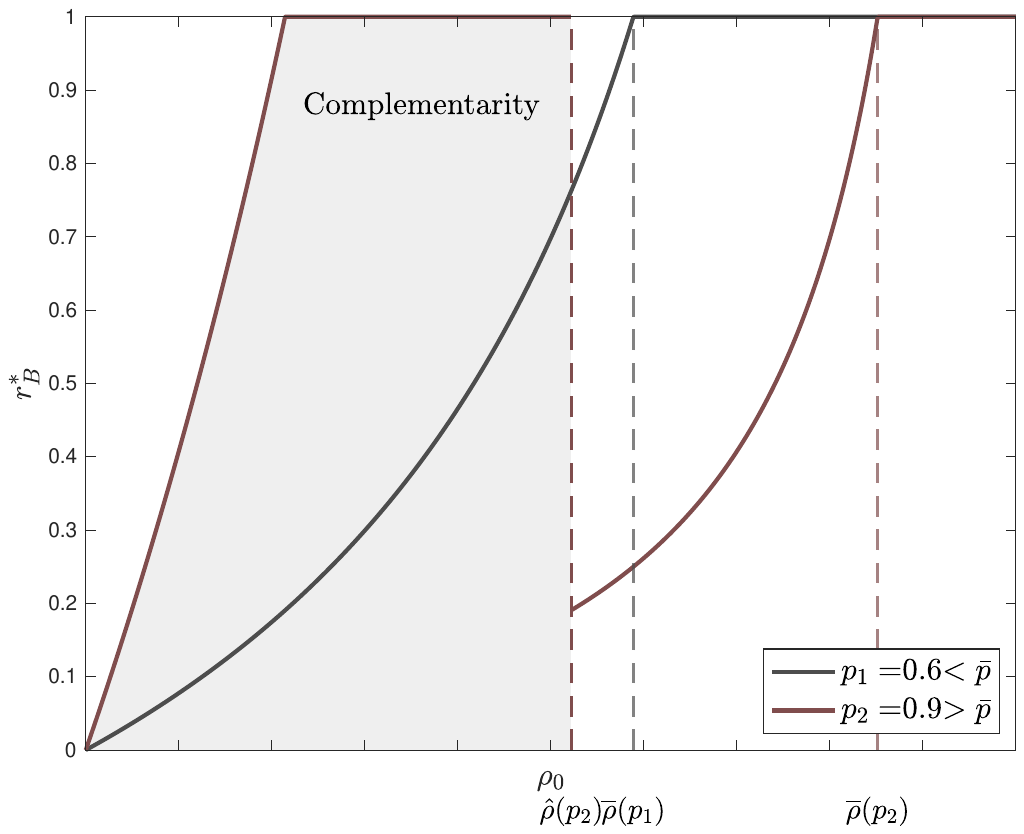}
        \caption{Sender's optimal strategy $r_B^*$ over $\rho_0\in[0,1]$}
        \label{fig:rB}
    \end{subfigure}%
    \hfill
    \begin{subfigure}[t]{0.48\textwidth}
        \centering
        \includegraphics[scale=0.4,trim=1cm 6.5cm 1.2cm 6.5cm,clip]{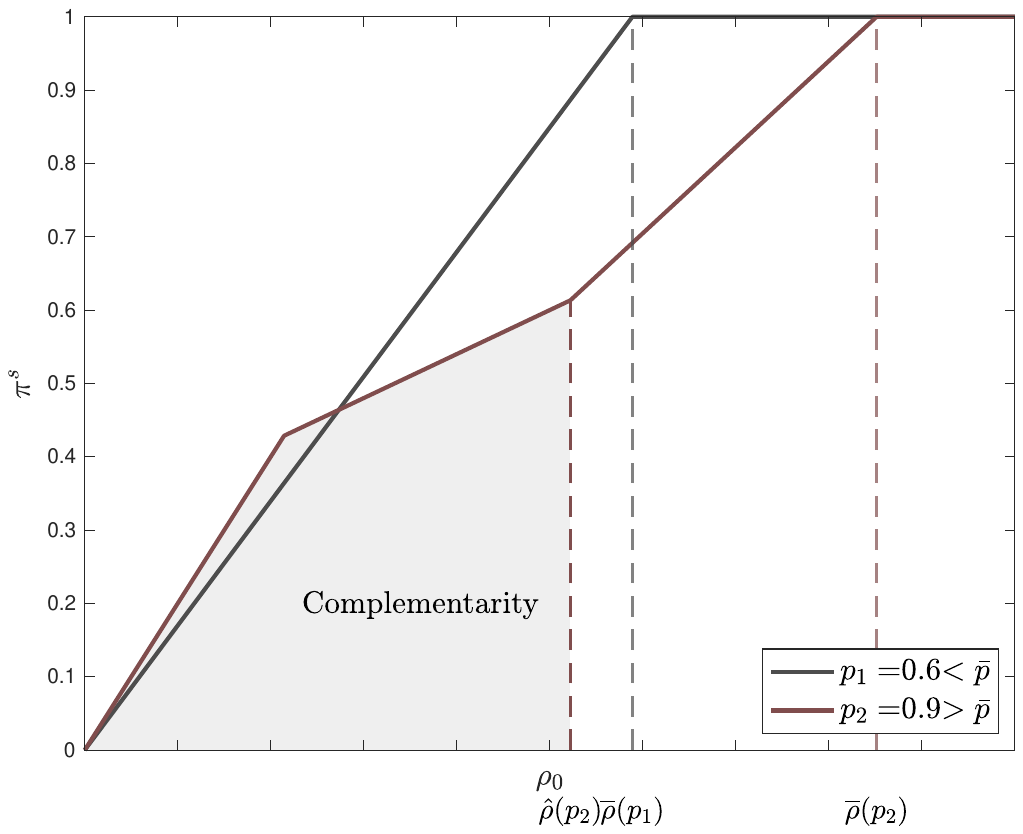}
        \caption{Sender's expected profit $\pi^s$ over $\rho_0\in[0,1]$}
        \label{fig:sender-profit}
    \end{subfigure}
    \caption{Comparative statics on sender's optimal strategy and profit under receiver internal  ($p=0.6$) vs. external ($p=0.9$) focus ($q = 0.3, v = 0.1$). The shaded region is where sender adopts a complementarity strategy.}
    \label{fig:strategy-vs-profit}
\end{figure}

Figure \ref{fig:eq-region-rho-vs-v} illustrates the equilibrium regions characterized by Propositions \ref{prop: auto} and \ref{prop: comp-and-self}. Figure \ref{fig:rB} illustrates the sender's optimal strategy under internal and external focus for a specific numerical example.\footnote{In Figure \ref{fig:rB}, we define $\overline{v}\equiv\frac{2 p q-3 p-q+2}{p-q}$ such that $p\leq \overline{p}\Leftrightarrow v\leq \overline{v}$. }

\medskip{}

We next characterize how the sender’s optimal strategy affects their expected profit across different regions of the parameter space.

\begin{proposition}\label{prop: profit}
(Sender's profit) 
The sender’s equilibrium expected profit for the three strategy regimes are as follows.  
\begin{itemize}[leftmargin=1em]
\item Automatic Affirmation: $\mathbf{E}_{\theta}\left(\pi^s\right)= 1$
\item Self-sufficiency:
$\mathbf{E}_{\theta}\left(\pi^s\right)= \rho_{0}\left(1+\frac{1 + v}{1 - v} \left(\frac{1-p}{1-q}\right)\right)<1$
\item Complementarity: $\mathbf{E}_{\theta}\left(\pi^s\right)= \min\left\{\rho_{0}p+(1-\rho_{0})q,\;\rho_{0}p\left(1+\frac{1+v}{1-v}\right)\right\}<1$
\end{itemize}
Moreover, $\frac{\partial\mathbf{E}_{\theta}\left(\pi^s\right)}{\partial p}>0$ if the sender adopts a complementarity strategy ($p\geq\overline{p}$, $\rho_0<\hat{\rho}$) and $\frac{\partial\mathbf{E}_{\theta}\left(\pi^s\right)}{\partial p}\leq 0$ otherwise. 
\end{proposition}

Proposition \ref{prop: profit} highlights that the sender benefits the most when the receiver automatically affirms the sender. In contrast, the sender's profit in the other two cases is lower because in each case the sender forgoes some opportunities to send inauthentic messages in order to preserve credibility. Depending on the parameters, the sender's profit under self-sufficiency could be higher or lower than that under complementarity. In other words, the higher value of $r_B^{\text{comp}}$ does not ensure a higher profit. 

Figure \ref{fig:sender-profit} illustrates the sender's profit under internal and external focus of the receiver for a specific numerical example.  Taken together, Figures \ref{fig:rB} and \ref{fig:sender-profit} show that lying more frequently through a higher $r^*_B$ does not necessarily mean a higher profit for the sender in equilibrium. The reason is that a higher $r_B^*$ does not always translate into a higher $\rho_2$, which is needed for a higher likelihood of the receiver supporting the sender. Holding all else fixed, when a higher $r^*_B$ is supported by a more favorable prior beliefs, i.e., a higher $\rho_0$, the sender makes a higher profit. Similarly, when stronger self-signaling incentives allow the sender to lie more often, the sender's profit increases.

Most interestingly, when the receiver is externally focused and the investigative technology improves, the sender earns a higher profit under the complementarity strategy. This is because the sender can free ride more on the investigator’s credibility, thereby gaining greater flexibility in persuading the receiver. This shows that when the external investigation contributes more in the coproduction of authenticity or inauthenticity, an improvement in the quality of external investigation helps the sender ($\frac{\partial\mathbf{E}_{\theta}\left(\pi^s\right)}{\partial p}> 0$).  As Proposition 3 shows, the reverse is true ($\frac{\partial\mathbf{E}_{\theta}\left(\pi^s\right)}{\partial p}\leq 0$) when the external investigation contributes less and the seller uses self-sufficiency strategy.   

\section{Receiver with Confirmation Bias} \label{sec: CB}

Just as the receiver's attitudes towards the specific pro-social cause, reflected in their self-signaling incentives, affect the sender's equilibrium strategies, the receiver's attitudes towards the sender could also play a significant role.  To examine this in more detail, we allow the receiver to overweight their prior belief about the sender in their updating process.  A large literature in psychology, economics and marketing (see for examples, Tversky and Kahneman 1974, Lord et al 1979, Grether 1980, Rabin and Schrag 1999, and Bond, et al. 2006) shows that people are not able to evaluate new information independent of their pre-existing initial information.  In particular, they tend to evaluate new information to be more consistent with their prior information.  In other words, the prior beliefs get a higher weight than in the classical Bayesian updating.  We incorporate this departure from Bayesian updating in this section.  We refer to this form of updating as confirmation bias because of the prevalence of that term, but we emphasize that there are other psychological phenomena (e.g., ``anchoring and adjustment" and ``pre-decisional distortion") that could also result in a decision-maker giving more weight to their prior belief than is prescribed by the Bayes rule.  Please note that we deviate from strict Bayesian updating but adhere to other parts of \cite{kamenica2011bayesian} framework. 

We model the receiver's confirmation bias by modifying the updating formula in \eqref{eq: rho1} as follows: 
\begin{equation}
   \rho_1(m=1) = 
\frac{k \rho_0 + (1-k) r_G \rho_0}{k \rho_0 + (1-k) r_G \rho_0 + k (1 - \rho_0) + (1 - k) r_B (1 - \rho_0)}
\label{eq: rho1-CB}
\end{equation}
and modifying the formula \eqref{eq: rho2} for updating after the investigative signal as follows:  
\begin{equation}
\rho_2(s) = 
\begin{cases} 
\frac{k\rho_1+(1-k)p \rho_1}{k\rho_1+(1-k)p \rho_1 +k(1-\rho_1)+(1-k)q (1 - \rho_1)} & \text{if } s = 1 \\[10pt]
\frac{k\rho_1+(1-k)(1 - p) \rho_1}{k\rho_1+(1-k)(1 - p) \rho_1 +k(1-\rho_1)+(1-k) (1 - q)(1 - \rho_1)} & \text{if } s = 0
\end{cases}
\label{eq: rho2-CB}
\end{equation}

In Equations \eqref{eq: rho1-CB} and \eqref{eq: rho2-CB}, the parameter $k \in [0,1]$ represents the extent of confirmation bias, with $k=0$ reducing to the case of no confirmation bias. In our formulation, the confirmation bias parameter, $k$, is multiplied with the relevant prior belief and added as an additional term.  The weight given to the new information is according reduced to $(1-k)$.{\footnote{The denominator is adjusted to appropriately scale the posterior probabilities.}} 

This formulation captures the behavioral pattern that confirmation bias causes receivers to overweight their prior beliefs relative to new evidence. Specifically, the parameter $k$ scales the influence of the prior in the updating process: when $k>0$, the receiver places additional weight on their prevailing prior belief ($\rho_0$ in Stage 2 or $\rho_1$ in Stage 3) relative to the likelihood ratios implied by the new information (firm's message or the investigator's signal). As a result, when the receiver has a more (less) favorable prior belief, they are more (less) likely to believe $m=1$ message or $s=1$ signal. When $k=0$, updating reduces to the standard Bayesian case; when $k=1$, the receiver ignores all new information and fully retains their prior belief.

In Stage 2, the receiver now overweights the prior beliefs in forming their posterior beliefs after observing the sender's message.  Therefore, regardless of the level of $r_B$ and $r_G$, $\rho_1(m=1)$, will be closer to $\rho_0$ with confirmation bias than without. In other words, both the positive effect of $r_G$ and the negative effect of $r_B$ on $\rho_{1}$ are dampened with confirmation bias. 
%Figure \ref{fig:isoposterior} illustrates three isoposterior curves in Stage 2 at varying levels of confirmation bias. 
In Stage 3, when the receiver sees the investigator's message, their confirmation bias leads to overweighting of $\rho_1$ and underweighting of the investigator's signal in developing the posterior beliefs, $\rho_2(m=1, s=1)$ and $\rho_2(m=1, s=0)$.  As a result, $\rho_2(m=1, s=1)$ and $\rho_2(m=1, s=0)$ will now be closer to $\rho_1$.  Essentially, the positive impact of the investigator's validation and the negative effect of their contradiction are both reduced. 

For the sender, conditional on the receiver supporting them, the benefits of receiving the support do not change with the receiver's confirmation bias.  However, for any set of $r_B, r_G$, and $s$, the receiver's posterior belief, $\rho_2$, and hence, the probability of the receiver supporting the sender will change. Therefore, our discussion here will focus on the effect of the changes in the belief updating process on our key results from Section 3.

In the previous section, Proposition \ref{prop: auto} showed that in the absence of confirmation bias, for a sufficiently high value of prior beliefs, $\rho_0>\overline{\rho}$, the sender took benefit of the receiver's automaticity and always sent a prosocial message.  We begin our analysis of the confirmation bias by examining how that strategy is affected by confirmation bias.  

\begin{proposition} \label{prop:affirm-reject}
{\textbf{(Automatic Affirmation and Rejection)}} 

There exists a level of prior belief $\overline{\overline{\rho}}(k) = \frac{\left(1-v\right)}{\left(1-v\right)+(1+v)\left(\frac{k+(1-k)(1-p)}{k+(1-k)(1-q)}\right)} <1$ such that for $\rho_0 \geq \overline{\overline{\rho}}(k)$, the sender always sends a prosocial message regardless of the state ($r_{G}^{*}=r_{B}^{*}=1$), and the receiver always supports them. In addition, there exists a value $\underline{\underline{\rho}}(k) \equiv \frac{\left(1-v\right)k}{\left(1-v\right)k+(1+v)\left(\frac{k+(1-k)p}{k+\left(1-k\right)q}\right)}<1$ such that, $\rho_0 <\underline{\underline{\rho}}(k)$, the receiver never supports the sender irrespective of the sender's strategy. 
\end{proposition}

Proposition \ref{prop:affirm-reject} shows that even with confirmation bias, if the receiver's prior belief is sufficiently favorable, i.e., $\rho_0>\overline{\overline{\rho}}$, the receiver will automatically support the sender. As the confirmation bias increases, the cutoff value $\overline{\overline{\rho}}$ decreases and the first part of Proposition \ref{prop:affirm-reject} is more likely to occur.  This might suggest that the receiver's confirmation bias results in them being more likely to support the sender's prosocial message.  Interestingly, however, that is not always the case.  When the receiver's prior belief is sufficiently unfavorable (pessimistic), i.e., $\rho_0 <\underline{\underline{\rho}}(k)$,  the receiver will not support the sender's reactive marketing message.  Further, a higher level of confirmation bias increases $\underline{\underline{\rho}}(k)$ and makes it less likely that the receiver will support the sender's prosocial message.{\footnote{Note that $\underline{\underline{\rho}}(k)=0$ when $k=0$ and $\rho_0 \nless \underline{\underline{\rho}}(k)$.}} Thus, confirmation bias operates as \emph{a double-edged sword}: A receiver with high prior beliefs is more susceptible to the sender's prosocial message, whereas a receiver with low prior beliefs becomes more impervious to the sender's message. 

The above discussion raises the question of what would happen when the prior belief is in the intermediate range, i.e., $\underline{\underline{\rho}}(k) \leq \rho_0 \leq \overline{\overline{\rho}}(k)$. Does the receiver's confirmation bias help or hurt the sender's reactive marketing strategy?  Table \ref{tab:rB-cB} in Appendix provides the values of $r_B^*$ across different parameter regions.

\begin{proposition}\label{prop: CB}
There exists a value of $p=\bar{\bar{p}}$ such that the sender adopts the self-sufficiency strategy when $p\leq\bar{\bar{p}}$. For $p>\bar{\bar{p}}$, the sender adopts either the self-sufficiency or complementarity strategy. Unlike in the baseline model, the threshold value $\bar{\bar{p}}$ increases in $\rho_0$, i.e., $\frac{\partial \bar{\bar{p}}}{ \partial \rho_0}>0$, due to confirmation bias.
\end{proposition}

Proposition \ref{prop: CB} shows that, as in the previous analysis, both complementarity and self-sufficiency strategies can arise in equilibrium. The sender's optimal choice between self-sufficiency or complementarity continues to depend on the relative emphasis in terms of internal and external focus.  However, the receiver's confirmation bias introduces a significant change. The cutoff value of $p=\bar{\bar{p}}$, now depends on the prior belief, $\rho_0$, and is an increasing function of $\rho_0$.  

How does the receiver's confirmation bias affect the equilibrium level of inauthenticity?  The answer to this question is complex because confirmation bias affects persuasion in two key ways. First, it undermines the persuasiveness of the sender’s prosocial message when the receiver’s prior belief is low, but enhances it when the prior belief is high. Second, it reduces the value of external investigation by weakening the positive effect of a validating signal and by mitigating the negative impact of a contradictory signal.

The next proposition describes the net impact of these changes on the values of $r_B$ that would be persuasive to the receiver in sender's self-sufficiency and complementarity strategy, respectively. 

\begin{proposition} (\textbf{Impact of Confirmation Bias on Inauthenticity})\label{prop: CB-impact}
    When the sender adopts self-sufficiency strategy ($r_B^*=r_B^{self}$), there exists a value of $\rho_0=\rho^+$ such that $\frac{\partial r_B^*}{\partial k}< (>0)$ for $\rho_0<(>)\rho^+$. When the sender adopts complementarity strategy ($r_B^*=r_B^{comp}$), we have $\frac{\partial r_B^*}{\partial k}\leq 0$.  
\end{proposition}

Proposition \ref{prop: CB-impact} shows that authenticity and inauthenticity continue to be coproduced but now with a greater contribution from the receiver and a smaller contribution from the investigator and the sender.  This is because the receiver's overall decision-making now places a lower emphasis on both the results on the external investigation and the sender's messaging strategy. We also find that the double-edged effect of confirmation bias continues with self-sufficiency strategy.  The extent of inauthenticity under self-sufficiency ($r_B^{self}$) could either increase or decrease with the extent of confirmation bias, depending on the prior beliefs ($\rho_0$). When the prior beliefs are lower, the sender needs to be more truthful to overcome confirmation bias; when the prior beliefs are higher, however, the sender can sustain more inauthentic messaging, as confirmation bias leads the receiver to interpret the message more favorably. In contrast, the extent of inauthenticity under complementarity, $r_B^{comp}$, decreases monotonically with confirmation bias. This occurs because confirmation bias reduces the value of the investigator’s validation.  In Stage 3, the receiver places greater weight on their belief $\rho_1$, which is increasingly shaped by their prior $\rho_0$ in Stage 2 due to confirmation bias. To achieve a sufficiently high $\rho_2$, the sender ends up being less inauthentic.

%\subsection{Impact of Self-Signaling and }

\section{Summary and Conclusions}
\label{sec: conclusion}

Businesses are a major part of our society and many consumers expect companies and their leaders to actively engage with the society and reflect prevailing societal values. In response, businesses often issue reactive marketing messages when an exogenous social event occurs, sometimes even when their businesses have loosely aligned with the underlying social cause. This raises an important tension: Consumers are known to dislike inauthentic prosocial behavior by businesses \citep[e.g.,][]{silver2021inauthenticity, karikari2024three} but they often lack full information about a company’s motives or true alignment with a cause. We examine this strategic question recognizing the fact that the interaction between businesses and consumers does not occur in isolation of other factors.  Specifically, we study three of the factors that affect businesses' and consumers' strategies and choices.  

\subsubsection*{External Investigation}We allow for the possibility that an external investigator can validate or contradict a reactive message's authenticity. Intuitively, one might expect that such fact-checking would discipline firms and curb inauthentic behavior. Our analysis shows that this intuition, while partly correct, is incomplete. Sometimes the sender leverages the investigator to enhance their own credibility (the complementarity strategy); other times they use a disciplined messaging (the self-sufficiency strategy) to establish credibility despite possible contradiction by the investigator. Moreover, it is possible that the presence of an external investigator actually increases inauthenticity in reactive messages.  Our results also show that as the investigator's signal becomes more precise, the receiver is more likely to have an external focus, which can lead to the sender using the complementarity strategy and a higher level of inauthenticity.   
% Last sentence: As q increases pbar goes down. For a given p, this makes complementarity more likely.  As p increases, p>pbar is more likely, again making complementarity more likely.  So the level of INauthenticity can go up.  This also ties better with the previous sentence.

\subsubsection*{Self-Signaling}

The receiver's self-signaling incentives have the potential to make consumers more open to pro-social messages. Our analysis shows that these incentives also enable the firm to engage in a higher level of inauthenticity in a variety of ways.  Specifically, a combination of strong self-signaling incentives and favorable prior beliefs lead to a zone of automatic affirmation where the sender does not need to ``prove'' its credibility to secure the receiver's support. In addition, as Proposition 2 shows, within self-sufficiency and complementarity strategies, self-signaling incentives of the receiver lead to a (weakly) higher-level of inauthenticity.  Our analysis also reveals a critical interaction: The impact of higher precision of the investigator's signal (mentioned above) is amplified as the receiver's self-signaling incentives increase.

We also note that the frequency of the sender's reactive messages, ($\rho_0 r_G^* + (1-\rho_0) r_B^*$), and the sender's profit increase as the receiver's self-signaling incentives become stronger.  

\subsubsection*{Confirmation Bias}

In contrast with the receiver's self-signaling incentives, their confirmation bias has a double-edged effect. Conceptually, the two demand-side mechanisms differ in nature. The self-signaling incentives directly expand the firm's \textit{capacity} for persuasion, raising the consumer's tolerance for some inauthenticity as long as the message itself aligns with their pro-social preference. In contrast, confirmation bias changes the \textit{credibility} of a message. It distorts how receivers process new information, affecting not only the perceived credibility of the firm’s messages but also the effectiveness of third-party investigation. When the prior beliefs are less favorable, the firm becomes more likely to switch from a self-sufficiency strategy to a complementarity strategy.  This is because under confirmation bias, the firm finds it difficult to maintain credibility on its own when the receiver underweights its claims.  This results in low firm profits even as it sends more reactive messages and a higher fraction of inauthentic messages. However, this pattern reverses when the priors are more favorable: The frequency of reactive and inauthentic messages might decline, but profit increases with confirmation bias. 

Our analysis also offers practical insight into when firms should engage in reactive marketing and when their messages are most likely to be perceived as credible. For instance, anchoring messaging to causes or events that consumers already strongly support can provide an additional cushion for establishing authenticity. At the same time, our model highlights that external investigation can influence not only how messages are perceived, but also how firms can design their reactive marketing to leverage external investigators.  On the flip side, consumers and regulators should be mindful that even well-intentioned mechanisms for transparency can be strategically incorporated into message design and hence cannot be universal panacea. Recognizing this dynamic is essential to understanding both the promise and limitations of credibility-enhancing tools in corporate messaging.

Overall, our work illustrates that the amount of reactive marketing messages and their authenticity arise from a complex interplay of psychological motivations, strategic behavior, and institutional design. Rather than being a simple matter of social engagement or ``truth versus spin,” reactive marketing equilibrium is coproduced by the sender, the receiver and the investigator. 

\subsubsection*{Limitations and Future Research}

To address realism concerns about the commitment assumption in standard Bayesian persuasion models, we can adopt a partial-commitment refinement in the spirit of \cite{lin2024credible}. In our setting, consumers observe the aggregate frequency of pro-social messages (vs. no action), analogous to the observed ``grade distributions'' in \cite{lin2024credible}. We can show that the firm’s ex ante commitment to a messaging strategy is \textit{credible} according to their definition, i.e., given how the receiver reacts to messages, the sender has no profitable deviation to any alternative disclosure policy that induces the same ex post message distribution (p. 2229). Here, commitment power does not arise from repeated interaction or traditional reputation mechanisms. Instead, the sender can exercise full commitment power because the distribution of their messages is publicly observable, making any deviation detectable at the aggregate level. In our model, the sender receives a fixed reward for receiver support regardless of the underlying state, so profit is maximized by maximizing the frequency of receiver's support, which the full-commitment Bayesian persuasion solution achieves exactly.\footnote{Further details and proofs are available upon request.} Consequently, the credibility constraint, as defined by \cite{lin2024credible}, imposes no additional cost on the sender relative to the standard full-commitment benchmark.

Although our analysis indicates that the sender could sometimes benefit from helping the investigator and sometimes from hindering the investigator, we have not derived an optimal level of obfuscation. When information about the cost of helping or hindering the investigator becomes better established, this analysis would become more relevant. Because our model incorporates multiple players, we have not been able to examine dynamic impact of reactive marketing strategies on the evolution of prior beliefs over time or any punishment by consumers.  Along similar lines, it is possible that for some controversial issues, some consumers might be against the issue whereas some others might be for it. Although such polarization adds an additional layer of complexity, our focus is less on polarization itself than on the firm’s credibility challenge: the pressure to respond to broadly appealing social causes and the temptation to appear authentic without genuine alignment. We hope that future research can address these questions.

\newpage
\bibliographystyle{apalike}
\bibliography{lit}

\newpage

\begin{appendix}

\input{appendix_survey}

\newpage
\input{appendix_extension}

\newpage
\input{appendix_proof}

\end{appendix}

\end{document}

%% file: appendix_survey.tex
\section*{Appendix A: Survey Results} 

Note: Numbers in each table are frequency percentages.  

\textbf{Question:} If a fact-checker (e.g., an individual fact-checker or an AI fact-checker like Grok on Twitter/X) \textbf{confirms} a pro-social message’s authenticity from a business leader or company, how much does that \textbf{increase} your likelihood to engage with it (e.g., like, share, comment)? 

\begin{table}[h!]
\centering
\begin{tabular}{ccccc}
\toprule
Not at all & A little & A moderate amount & A lot & A great deal \\
\midrule
13.5 & 24.5 & 28.5 & 21.5 & 12 \\
\bottomrule
\end{tabular}
\end{table}
    
\vspace{1em}

\noindent
\textbf{Question:} If a fact-checker (e.g., an individual fact-checker or an AI fact-checkers like Grok on Twitter/X) \textbf{disputes} a pro-social message’s authenticity, how much does that \textbf{decrease} your likelihood to engage with it (e.g., like, share, comment)?

\begin{table}[h!]
\centering
\begin{tabular}{ccccc}
\toprule
Not at all & A little & A moderate amount & A lot & A great deal \\
\midrule
12 & 17.5 & 25.5 & 22.5 & 22.5 \\
\bottomrule
\end{tabular}
\end{table}

\vspace{1em}

\textbf{Question:}  Would you support a pro-social message in the following cases, even if you do not know (cannot figure out) if it is genuine?

When you strongly believe in the cause promoted in the message:

\begin{table}[h!]
\centering
\begin{tabular}{>{\centering\arraybackslash}p{0.18\textwidth}
                >{\centering\arraybackslash}p{0.18\textwidth}
                >{\centering\arraybackslash}p{0.2\textwidth}
                >{\centering\arraybackslash}p{0.18\textwidth}
                >{\centering\arraybackslash}p{0.18\textwidth}}
\toprule
Strongly disagree & Somewhat disagree & Neither agree nor disagree & Somewhat agree & Strongly agree \\
\midrule
8 & 8 & 13.5 & 47 & 23.5 \\
\bottomrule
\end{tabular}
\end{table}

When you strongly admire the business leader or the company that sent the message:

\begin{table}[h!]
\centering
\begin{tabular}{>{\centering\arraybackslash}p{0.18\textwidth}
                >{\centering\arraybackslash}p{0.18\textwidth}
                >{\centering\arraybackslash}p{0.2\textwidth}
                >{\centering\arraybackslash}p{0.18\textwidth}
                >{\centering\arraybackslash}p{0.18\textwidth}}
\toprule
Strongly disagree & Somewhat disagree & Neither agree nor disagree & Somewhat agree & Strongly agree \\
\midrule
10 & 13.5 & 24 & 35.5 & 17 \\
\bottomrule
\end{tabular}
\end{table}

\vspace{2em}

\textbf{Question:} When you engage with a pro-social message from a business leader or company (e.g., like, share, comment), does it make you feel good about your own values, beliefs, or identity?

\begin{table}[h!]
\centering
\begin{tabular}{ccccc}
\toprule
Not at all & A little & A moderate amount & A lot & A great deal \\
\midrule
13.4 & 31.8 & 27.4 & 19.4 & 8 \\
\bottomrule
\end{tabular}
\end{table}

\vspace{1em}

\textbf{Question:}  Does supporting pro-social messages from business leaders or companies reinforce your belief that you are a person who cares about social causes?

\begin{table}[h!]
\centering
\begin{tabular}{ccccc}
\toprule
Not at all & A little & A moderate amount & A lot & A great deal \\
\midrule
18.9 & 27.4 & 26.4 & 18.4 & 9 \\
\bottomrule
\end{tabular}
\end{table}

%% file: appendix_extension.tex
\section*{Appendix B: Multiple Receiver Groups} 
\label{sec: ext}

In this section, we analyze a model that includes multiple groups of receivers who are \textit{ex-ante} identical as in the main model but are different \textit{ex-post} after being exposed to different information sets. Specifically, we consider a scenario where, \textit{ex post}, there are three groups of receivers: (i) Those who observe only the sender's message $m$, (ii) Those who observe both the sender's message $m$ and the investigator's signal $s$, and (iii) Those who observe neither.\footnote{In our framework, a hypothetical group of receivers observing only the investigator's signal is not feasible, as the signal is designed to assess the veracity of the sender's message and must, therefore, accompany it.} We define these receiver groups/types ($g$) as follows: 
\begin{itemize}

\item \textit{Actives} ($g = MS$): These receivers observe both the sender’s message $m$ and the investigator’s signal $s$. They are likely more engaged and discerning consumers of information, who not only encounter the sender's claims but also actively seek verification—such as by reading fact-checking articles, following investigative journalists, or consuming posts from trusted experts. They might also be exposed to the signal by the platform's algorithm. 

\item \textit{Occasionals} ($g = M$): This group consists of receivers who observe only the sender's message $m$ and no additional investigation signal. They represent casual social media users who encounter information through algorithmic feeds or trending topics.\footnote{ They might not be shown the investigator's signal by the platform's algorithm and/or they do not actively seek out investigator's signal. We are agnostic about the reason for them not observing the signal.} 

\item \textit{Uninformed} ($g = N$): This group does not receive either the sender's message or the investigator’s signal, making them entirely uninformed about the issues at hand. 

\end{itemize}

We let $g \in \{M, MS, N\}$ denote the receiver group types and define their population shares as $\alpha_M \geq 0$, $\alpha_{MS} \geq 0$, and $\alpha_N \geq 0$, corresponding to the fractions of receivers in each category. We impose the constraint $\sum_{g\in\{M,MS,N\}}\alpha_{g}=1$ to ensure well-defined population shares. 

The sender's objective function $\pi^s$ is now given by the weighted sum of payoff obtained from each segments of receivers, we assume that 
\[\pi^s =\sum_{g\in\{M,MS,N\}}\alpha_{g}\cdot \mathbf{1}(a_g) \cdot \underline{\pi},\] where $\underline{\pi}>0$ is the unspecified utility from getting a positive brand image, and $\mathbf{1}(a_g)=1$ if $a_g=1$, i.e., the receiver of group $g$ supports the sender, and $0$ otherwise. 

Note that our main model is a special case of this generalized framework with multiple receiver segments, corresponding to the scenario where $\alpha_{MS} = 1$. By introducing multiple receiver segments, this framework allows us to investigate how the heterogeneity in receivers' information-seeking behavior affects equilibrium outcomes. 

Borrowing the notation from the benchmark model, one can easily verify the following response function from each group of receivers: 
\[\mathbf{1}(a_{g})=\begin{cases}
0 & \text{if }g=N\\
\mathbf{1}\left(\rho_{1}>\frac{1}{2}\left(1-v\right)\right) & \text{if }g=M\\
\mathbf{1}\left(\rho_{2}>\frac{1}{2}\left(1-v\right)\right) & \text{if }g=MS
\end{cases}.\]
We use $(r_G^*,r_B^*)$ and $(r_G^{MR},r_B^{MR})$ respectively to denote the sender's equilibrium strategy in the benchmark model and in this extension section. It is straightforward to show that $r_G^{MR}=r_G^*=1$. 

\begin{proposition}\label{prop: multi}
When the benchmark strategy is complementarity (i.e., $p > \overline{p}$ and $\rho_0 < \hat{\rho}$), the sender is authentic more frequently in the multi-receiver setting, i.e., $r_B^{MR} < r_B^*$, if a large enough proportion of receivers observe only the sender's message: 
\[\frac{\alpha_M}{\alpha_{MS}} >
\min\left\{
\frac{1}{2}(1 + v)(p - q), \,
\frac{2p(1 - q)}{2 - p(1 + v) - q(1 - v)} - 1
\right\}.\]
When the benchmark strategy is self-sufficiency (i.e., $p \leq \overline{p}$ or $\hat{\rho} < \rho_0 < \overline{\rho}$), the sender is authentic less frequently in the multi-receiver setting, i.e., $r_B^{MR} > r_B^*$, if  a sufficiently large proportion of receivers observe only the sender’s message:
\[\frac{\alpha_M}{\alpha_{MS}} > \frac{(1-v)(1-p)(1-q)+(1+v)(1-p-q+q^2)}{(1+v)(p-q)}.\]

Otherwise, the sender maintains the benchmark strategy, i.e., $r_B^{MR} = r_B^*$.
\end{proposition}

When the share of ``occasionals,'' those who observe only sender's messages, is sufficiently large, the sender can be paradoxically \textit{more} often authentic. This occurs because the sender gains greater strategic flexibility so that they can shift their focus from convincing the ``actives'' by leveraging the positive validation of the investigation to persuading the ``occasionals'' through increasing the persuasiveness of their direct message.  Said differently, the sender is unable to leverage the external investigators with the ``occasionals" segment, which increases the gains from a higher level of authenticity for the sender.  

The above analysis shows that the main insights from the single receiver model apply in the current case but the relative size of the ``occassionals" can shift the specific solution one way or the other.

%% file: appendix_proof.tex
\section*{Appendix C: Proofs}
\paragraph{Proof of Lemma \ref{lem: posterior}}

%We prove this Lemma by contradiction. 

First, in any equilibrium, we must have $r_G\geq r_B$.  

Suppose, for contradiction, that in equilibrium $r_G^*<r_B^*$. The receivers' posterior belief ($\rho_2$) upon observing a prosocial message ($m=1$) increases in $r_G$ whereas decreases in $r_B$. Since $r_G^*<r_B^*$, sending prosocial message is a weaker signal of authenticity, reducing $\rho_2$. The sender can strictly improve their payoff by deviating to $\hat{r_G}=r_B^*+\varepsilon$, for some $\varepsilon>0$. This deviation (1) increases $\rho_2$ since $\hat{r_G}>r_B^*$, making prosocial claims more credible, and (2) raises the frequency of $m=1$, improving the sender's expected utility. This contradicts the assumption that $(r_G^*,r_B^*)$ is an equilibrium.

Next, we show that $\rho_{2}\left(1,s\right)\geq\rho\left(s\right)$ given that $r_{G}\geq r_{B}$.
\begin{align*}
\rho_{2}\left(1,s\right)	&=\frac{Pr\left(\theta=1,m=1|s\right)}{Pr\left(m=1|s\right)} \\
	& =\frac{Pr\left(m=1|\theta=1\right)\cdot Pr\left(\theta=1|s\right)}{Pr\left(m=1|\theta=0\right)\cdot Pr\left(\theta=0|s\right)+Pr\left(m=1|\theta=1\right)\cdot Pr\left(\theta=1|s\right)} \\
	& =\frac{r_{G}\cdot Pr\left(\theta=1|s\right)}{r_{B}\cdot Pr\left(\theta=0|s\right)+r_{G}\cdot Pr\left(\theta=1|s\right)}.
\end{align*}
Dividing both the denominator and numerator by $r_{G}>0$, we obtain 
\begin{align*} 
\rho_{2}\left(1,s\right)	& =\frac{Pr\left(\theta=1|s\right)}{\frac{r_{B}}{r_{G}}\cdot Pr\left(\theta=0|s\right)+Pr\left(\theta=1|s\right)} \\ 
	& \geq\frac{Pr\left(\theta=1|s\right)}{Pr\left(\theta=0|s\right)+Pr\left(\theta=1|s\right)}
	=\rho\left(s\right), 
\end{align*}
where the inequality follows from $r_{G}\geq r_{B}$. 
% Now suppose we also have $\rho_{2}\left(0,s\right)>\rho\left(s\right)$. Then we can write the expected posterior belief as below:
% \begin{align*}
% \mathbf{E}_{m}\left[\rho_{2}\left(m,s\right)\right] &=	Pr\left(m=0|s\right)\cdot\rho_{2}\left(0,s\right)+Pr\left(m=1|s\right)\cdot\rho_{2}\left(1,s\right) \\
% &>	Pr\left(m=0|s\right)\cdot\rho\left(s\right)+Pr\left(m=1|s\right)\cdot\rho\left(s\right) \\
% & =	\rho\left(s\right).
% \end{align*}
% This contradicts the law of iterated expectation, given by equation \eqref{eq: martingale}. 
\qed
\bigskip{}

\paragraph{Proof of Proposition \ref{prop: auto}}

Recall that in stage 4, the receiver will support the sender's prosocial claim if and only if their posterior belief about the sender being a good type exceeds a threshold, i.e., $\rho_2(m,s)>\frac{1}{2}(1-v)$.

Define $\rho(s)$ as the updated belief from observing only the signal $s$. Since the investigation signal is informative ($1>p>\frac{1}{2}>q>0$), it follows that $\rho (s=1)>\rho(s=0)$. 
According to Lemma \ref{lem: posterior} and the martingale property of Bayes update, if the belief after observing $s = 0$ satisfies:
\begin{align}
&\rho(s=0)=\frac{\rho_{0}\left(1-p\right)}{\rho_{0}\left(1-p\right)+\left(1-\rho_{0}\right)\left(1-q\right)}>\frac{1}{2}(1-v), \label{eq:rhos0} 
\end{align}
then it must be that 
\begin{equation}
\rho_2(m=1,s=1)>\rho_2(m=1,s=0)\geq \rho(s=0)>\frac{1}{2}(1-v).
\end{equation}
It is easy to verify that \eqref{eq:rhos0} can be rewritten as the condition below: \[\rho_0>\overline{\rho}\equiv\frac{\left(1-v\right)\left(1-q\right)}{\left(1-v\right)\left(1-q\right)+\left(1+v\right)\left(1-p\right)}.\] 
That is, the receiver will support the sender's prosocial message regardless of the realized signal $s$ if $\rho_0>\overline{\rho}$. Anticipating this outcome, the sender always sends the prosocial message regardless their type, i.e., $r_G^*=r_B^*=1$.

Next we show that,  if $\rho_0\leq\overline{\rho}$, or equivalently, if $\rho(s=0)\leq \frac{1}{2}(1-v)$, always sending prosocial message ($r_G^*=r_B^*=1$) is a (weakly) dominated strategy. 

Suppose the sender chooses $r_G^* = r_B^* = 1$, i.e., always sends the message $m=1$ regardless of the state. Then the message carries no information, and the posterior beliefs upon observing $m=1$ will equal the updated belief from observing the signal alone, that is, we have:
\[\rho_2(m=1, s=0)  = \rho(s=0) \leq \frac{1}{2}(1 - v).\]
Hence, receivers do not support the sender after observing $(m=1, s=0)$, and the sender receives a payoff of zero in that case.
Recall that the sender's objective function is given by 
\[ 
\pi\left(r_{G},r_{B}\right)=\sum_{s\in\left\{ 0,1\right\} }Pr\left(m=1,s\right)\cdot1\left[\mathbf{\rho_{2}}\left(m=1,s\right)\geq\frac{1}{2}\left(1-v\right)\right],\]
where the probability of prosocial message $Pr\left(m=1,s\right)$ strictly increases in both $r_G$ and $r_B$, and the receivers' posterior belief $\mathbf{\rho_{2}}\left(m=1,s\right)$ increases in $r_G$ but decreases in $r_B$. 

Therefore, the sender's optimal strategy is to always set $r_G^*=1$ and choose between two values for $r_B^*$, corresponding to one of the following strategies:
\begin{itemize}
\item A \textit{self-sufficiency} strategy, with $r_B^*=r_B^{\text{self}}$, which secures support in both signal realizations by ensuring $\rho_2(m=1, s=1)>\rho_2(m=1, s=0) \geq \frac{1}{2}(1-v)$; or  
\item A \textit{complementarity strategy}, with $r_B^*=r_B^{\text{comp}}$, which exploits the signal's validation by inducing support only when $s=1$, i.e., $\rho_2(m=1, s=1) \geq \frac{1}{2}(1-v)>\rho_2(m=1, s=0)$,
\end{itemize}
where $0<r_B^{\text{self}} < r_B^{\text{comp}}\leq 1$. 

The next proposition characterizes the conditions under which each strategy is optimal and thus completes the proof of the sender's equilibrium behavior.
\qed
\bigskip{}

\paragraph{Proof of Proposition \ref{prop: comp-and-self}}

We proceed under the assumption that $\rho_0 < \overline{\rho}$, or equivalently, that the posterior after observing $s = 0$ satisfies: $\rho(s=0) \leq \frac{1}{2}(1 - v)$.

We first characterize the self-sufficiency and complementarity strategy, respectively,  and then show that only the former is optimal under weak investigative technology, i.e., $p \leq \overline{p} = \frac{2 - (1 - v)q}{3 - 2q + v}$. 

According to Bayes' rule, the posterior belief after $(m=1, s=0)$ is:
\[\rho_2(m=1, s=0) = \frac{\rho_0(1 - p)r_G}{\rho_0(1 - p)r_G + (1 - \rho_0)(1 - q)r_B}.\]
Given that $r_G^*=1$, solving $\rho_2(m=1, s=0) \geq \frac{1}{2}(1 - v)$ yields:
\[r_B \leq \left( \frac{1 - p}{1 - q} \right) \left( \frac{1 + v}{1 - v} \right) \left( \frac{\rho_0}{1 - \rho_0} \right).\]
This defines the optimal $r_B^{\text{self}}$ in the self-sufficiency strategy:
\[r_B^{\text{self}} = \left( \frac{1 - p}{1 - q} \right) \left( \frac{1 + v}{1 - v} \right) \left( \frac{\rho_0}{1 - \rho_0} \right).\]
It is easy to verify that the self-sufficiency strategy is always feasible ($r_B^{\text{self}} \leq 1$) when $\rho_0 < \overline{\rho}$. 
With this strategy, the sender's payoff is given by 
\begin{equation}\label{eq: pi-self}
\pi^s_{\text{self}}:= \mathbb{E}_\theta[\pi^s(r_G^* =1, r_B^* = r_B^{\text{self}})] = \rho_{0}\left(1+\left( \frac{1 + v}{1 - v} \right) \cdot\frac{1-p}{1-q}\right).
\end{equation}

Next we characterize the complementarity strategy where receivers support only if they observe $m=s=1$.

According to Bayes' rules, the posterior belief after observing $(m=1, s=1)$ is:
\[\mathbf{\rho_{2}}\left(m=1,s=1\right)=\frac{pr_{G}\rho_{0}}{q\left(1-\rho_{0}\right)r_{B}+pr_{G}\rho_{0}}.\]
Solving for the maximum (and feasible) $r_B^{\text{comp}}$ that ensures $\rho_2(m=1, s=1) \geq \frac{1}{2}(1-v)$ yields:
\[r_{B}^{\text{comp}}=\begin{cases}
\left(\frac{p}{q}\right)\left(\frac{1+v}{1-v}\right)\left(\frac{\rho_{0}}{1-\rho_{0}}\right) & \text{if }\rho_{0}<\underline{\rho}\equiv\frac{\left(1-v\right)\left(1-p\right)}{\left(1-v\right)\left(1-p\right)+p\left(1+v\right)}\\
1 & \text{if }\rho_{0}\geq\underline{\rho}
\end{cases}.\]
 With this strategy, the sender's payoff is given by 
\begin{equation}\label{eq: pi-comp}
\pi^s_{\text{comp}}:= \mathbb{E}_\theta[\pi^s(r_G^* =1, r_B^* = r_B^{\text{comp}})] 
=\begin{cases}
\rho_{0}p\left(1+\frac{1+v}{1-v}\right) & \text{if }\rho_{0}<\underline{\rho}\\
\rho_0 p + (1-\rho_0) q & \text{if }\rho_{0}\geq\underline{\rho}
\end{cases}.
\end{equation}
Hence the self-sufficiency strategy is the optimal strategy, i.e. $r_B^*=r_B^{\text{self}}$, if $\pi^s_{\text{self}}\geq\pi^s_{\text{comp}}$. Note that $\rho_{0}p\left(1+\frac{1+v}{1-v}\right)\geq \rho_0 p + (1-\rho_0) q$ when $\rho_0\geq\underline{\rho}$.  Thus, a sufficient condition for \eqref{eq: pi-self} $\geq$ \eqref{eq: pi-comp} can be re-written as:
\begin{equation}
1+\left(\frac{1+v}{1-v}\right)\cdot\frac{1-p}{1-q} \geq p\left(1+\frac{1+v}{1-v}\right)
\end{equation}
Solving this inequality yields the upper bound $\overline{p}$ as a function of $(v, q)$. Proposition \ref{prop: comp-and-self} (i) follows from identifying the threshold $\overline{p}$:
\[p \leq \overline{p} = \frac{2 - (1 - v)q}{3 - 2q + v}.\]

\medskip{}
Next we prove the results stated in Proposition \ref{prop: comp-and-self} (ii) when $p>\overline{p}= \frac{2 - (1 - v)q}{3 - 2q + v}$. 

% As shown above, the self-sufficiency strategy secures receivers' support in both signal realizations by setting $r_G^* = 1$ and choosing:
% \[r_B^* = r_B^{\text{self}} = \left( \frac{1 - p}{1 - q} \right) \left( \frac{1 + v}{1 - v} \right) \left( \frac{\rho_0}{1 - \rho_0} \right),\]
% yielding sender payoff:
% \begin{equation}\label{eq:pi-self-2}
% \pi^s_{\text{self}} = \rho_0 \left( 1 + \left( \frac{1 + v}{1 - v} \right) \cdot \frac{1 - p}{1 - q} \right).
% \end{equation}

% The complementarity strategy, by contrast, induces support only when $m = s = 1$, and requires:
% \[r_B^* = r_B^{\text{comp}} = \min\left\{1,\left( \frac{p}{q} \right) \left( \frac{1 + v}{1 - v} \right) \left( \frac{\rho_0}{1 - \rho_0} \right)\right\},\]
% with resulting payoff:
% \begin{equation}\label{eq:pi-comp-2}
% \pi^s_{\text{comp}} = \begin{cases}
% \rho_{0}p\left(1+\frac{1+v}{1-v}\right) & \text{if }\rho_{0}<\underline{\rho}\equiv\frac{\left(1-v\right)\left(1-p\right)}{\left(1-v\right)\left(1-p\right)+p\left(1+v\right)}\\
% \rho_{0}p+(1-\rho_{0})q & \text{if \ensuremath{\rho_{0}}}\geq\underline{\rho}
% \end{cases}.
% \end{equation}

The sender compares the two strategies and prefers complementarity if: $\pi^s_{\text{comp}} > \pi^s_{\text{self}}$. 
In the case of $\rho_0<\underline{\rho}$, as shown above, this condition is equivalent to $p>\overline{p}$.

In the case of $\rho_0\geq \underline{\rho}$, this condition is equivalent to:
\begin{equation}
\rho_{0}p+(1-\rho_{0})q>\rho_{0}\left(1+\left(\frac{1+v}{1-v}\right)\cdot\frac{1-p}{1-q}\right).\label{eq: cond1}  
\end{equation}
Re-arranging condition \eqref{eq: cond1} yields $\rho_0<\hat{\rho}\equiv \frac{(1-q) q (1-v)}{p (q (1-v)-2)+q^2 (v-1)+2}$.

Thus, for $p > \overline{p}$, the sender strictly prefers the complementarity strategy when $\rho_0<\hat{\rho}$ since it yields a higher payoff than the self-sufficiency strategy. In this region, the sender sets:  
\[r_B^*=r_B^{\text{comp}}=\min\left\{1,\left( \frac{p}{q} \right) \left( \frac{1 + v}{1 - v} \right) \left( \frac{\rho_0}{1 - \rho_0} \right)\right\}.\]

When $\hat{\rho} \leq \rho_0 < \overline{\rho}$, the self-sufficiency strategy becomes payoff-dominant, and the sender optimally switches to:
\[r_B^* = r_B^{\text{self}}=\left( \frac{1 - p}{1 - q} \right) \left( \frac{1 + v}{1 - v} \right) \left( \frac{\rho_0}{1 - \rho_0} \right).\]

This completes the proof of Proposition \ref{prop: comp-and-self}.
\qed

\bigskip{}

\paragraph{Proof of Proposition \ref{prop: profit}}
The sender's profit function follows directly from the characterizations in Proposition \ref{prop: comp-and-self}, which specify the sender's best response and resulting payoff in each regime. To summarize, we have
\[\mathbb{E}_{\theta}\left(\pi^{s}\right)=\begin{cases}
1 & \text{if }\rho_{0}>\overline{\rho}\\
\pi_{\text{self}}^{s}=\rho_{0}\left(1+\left(\frac{1+v}{1-v}\right)\cdot\frac{1-p}{1-q}\right) & \text{if }\rho_{0}<\overline{\rho},p<\overline{p}\text{ or }\hat{\rho}<\rho_{0}\leq\overline{\rho},p\geq\overline{p}\\
\pi_{\text{comp}}^{s}=\rho_{0}p+(1-\rho_{0})q & \text{if }\underline{\rho}<\rho_{0}\leq\hat{\rho},p\geq\overline{p}\\
\pi_{\text{comp}}^{s}=\rho_{0}p\left(1+\frac{1+v}{1-v}\right) & \text{if }\rho_{0}\leq\underline{\rho},p\geq\overline{p}
\end{cases}.\] 
\qed

\bigskip{}

\paragraph{Proof of Proposition \ref{prop:affirm-reject}}
There exists $\overline{\overline{\rho}}(k) = \frac{\left(1-v\right)}{\left(1-v\right)+(1+v)\left(\frac{k+(1-k)(1-p)}{k+(1-k)(1-q)}\right)} <1$ such that for $\rho_0 \geq \overline{\overline{\rho}}(k)$, the sender always sends a prosocial message regardless of the state ($r_{G}^{*}=r_{B}^{*}=1$), and the receiver always support them. In addition, there exists a value $\underline{\underline{\rho}}(k) \equiv \frac{\left(1-v\right)k\left[k+\left(1-k\right)q\right]}{\left(1-v\right)k\left[k+\left(1-k\right)q\right]+(1+v)\left[k+(1-k)p\right]}<1$ such that, for $\rho_0 <\underline{\underline{\rho}}(k)$, the receiver never supports the sender irrespective of the sender's strategy.

We analyze the sender’s strategy under confirmation bias. The receiver updates their belief in two stages, first upon observing the sender’s message $m$, and then upon observing the signal $s$. 

Recall that confirmation bias is modeled by attenuating the weight placed on new information, controlled by parameter $k \in (0,1]$. When $k=1$, the receiver is fully Bayesian; as $k \to 0$, beliefs become increasingly dependent on prior.

The receiver will support the sender if and only if the posterior belief in stage 4, $\rho_2(m,s)$, exceeds the threshold $\frac{1}{2}(1-v)$. We seek thresholds on the prior $\rho_0$ that determine when the sender is guaranteed to receive support (regardless of $s$) or never receives support (under any strategy).

1. Threshold for Automatic Affirmation:

Suppose the sender always sends the prosocial message: $r_G = r_B = 1$. Then the message is uninformative and $\rho_1(m=1) = \rho_0$. The second-stage update from the signal $s=0$ is thus given by formula \eqref{eq: rho2-CB}:
\[\rho_2(s=0) = \frac{k\rho_0+(1-k)(1 - p) \rho_0}{k\rho_0+(1-k)(1 - p) \rho_0 +k(1-\rho_0)+(1-k) (1 - q)(1 - \rho_0)}\]
We require:
\[\rho_2(s=0) \geq \frac{1}{2}(1 - v),\]
to ensure support even in the worst-case signal realization ($s=0$).

Solving this inequality yields the upper threshold:
\[\rho_0 > \overline{\overline{\rho}}(k) \equiv \frac{(1 - v)[k + (1 - k)(1 - q)]}{(1 - v)[k + (1 - k)(1 - q)] + (1 + v)[k + (1 - k)(1 - p)]}.\]

%After simplification, this expression can be written as:\[\overline{\overline{\rho}}(k) = \frac{(v - 1)[(k - 1)q + 1]}{-(k - 1)p(v + 1) + (k - 1)q(v - 1) - 2}.\]

Hence, for $\rho_0 > \overline{\overline{\rho}}(k)$, the receiver supports the sender regardless of $s$, and the sender optimally chooses $r_G^* = r_B^* = 1$.

2. Threshold for Automatic Rejection:

To identify when the sender never receives support under any strategy, we consider the most optimistic case for the sender: when the message is maximally informative. This occurs under a fully separating strategy, $r_G = 1$, $r_B = 0$, in which only the good type sends the prosocial message. Under this strategy, the receiver’s posterior $\rho_2(m=1, s=1)$ is maximized. If even this highest possible posterior fails to exceed the threshold $\frac{1}{2}(1 - v)$, then no strategy can elicit support.

The posterior after observing the message is given by:
\[\rho_1(m=1) = \frac{\rho_0}{\rho_0 + (1 - \rho_0)k}.\]
Due to confirmation bias ($k>0$), this posterior remains strictly less than 1 even if the sender is always authentic ($r_G=1,r_B=0$). 

The posterior belief after observing $s=1$ is:
\[\rho_2(s=1) = \frac{\rho_1[k + (1 - k)p]}{\rho_1[k + (1 - k)p] + (1 - \rho_1)[k + (1 - k)q]}.\]

Substituting $\rho_1$ and simplifying yields an expression in terms of $\rho_0$:
\[\rho_2(s=1) = \frac{\rho_{0}[k+(1-k)p]}{\rho_{0}[k+(1-k)p]+(1-\rho_{0})k[k+(1-k)q]}.\]

We require:
\[\rho_2(s=1) < \frac{1}{2}(1 - v),\]
which after solving gives the lower threshold:
\[\rho_0 < \underline{\underline{\rho}}(k) \equiv \frac{\left(1-v\right)k\left[k+\left(1-k\right)q\right]}{\left(1-v\right)k\left[k+\left(1-k\right)q\right]+(1+v)\left[k+(1-k)p\right]}.\]

Thus, if $\rho_0 \leq \underline{\underline{\rho}}(k)$, even the highest possible posterior falls below the threshold, and the receiver never supports the sender regardless of their strategy.
\qed

\bigskip{}

\paragraph{Proof of Proposition \ref{prop: CB}}

For $\underline{\underline{\rho}}(k) \leq \rho_0 < \overline{\overline{\rho}}(k)$, as in the benchmark model, the sender chooses between two strategies: (i) self-sufficiency: $r_B = r_B^{\text{self}}$, ensuring support under both $s = 0$ and $s = 1$; or (ii) complementarity: $r_B = r_B^{\text{comp}}$, inducing support only when $s = 1$.

The posterior belief after observing $(m=1, s=0)$ is:
\begin{align*}
&\rho_2(m=1, s=0) \\
= &\frac{[k + (1 - k)(1 - p)] \rho_1(m=1)}{[k + (1 - k)(1 - p)] \rho_1(m=1) + [k + (1 - k)(1 - q)] (1 - \rho_1(m=1))},
\end{align*}
where
\[\rho_1(m=1) = \frac{\rho_0}{\rho_0 + k(1 - \rho_0) + (1-k)(1-\rho_0)r_B }.\]
Substituting into $\rho_2(m=1, s=0) \geq \frac{1}{2}(1 - v)$ and solving for $r_B$ gives the upper bound:
\[r_B^{\text{self}} = \frac{\left(\frac{1+v}{1-v}\right)\left(\frac{\rho_{0}}{1-\rho_{0}}\right)\left(\frac{1-\left(1-k\right)p}{1-\left(1-k\right)q}\right)-k}{1-k}.\]

This is the maximum $r_B$ allowed for the sender adopting self-sufficiency to secure support from receivers with confirmation bias. It is easy to verify that when $\underline{\underline{\rho}}(k) \leq \rho_0 < \overline{\overline{\rho}}(k)$, we must have $r_B^{\text{self}}\leq1$, for any $k\in[0,1]$. However, self-sufficiency is feasible, i.e., $r_B^{\text{self}}\geq0$, only if $p \leq p_1$, where 
\[p_1\equiv \frac{\rho _0 (k (v-1) ((k-1) q+1)-v-1)-k (v-1) ((k-1) q+1)}{(k-1) \rho _0 (v+1)}.\]

Similarly, the posterior belief after observing $(m=1, s=1)$ is:
\[\rho_2(m=1, s=1) = \frac{[k + (1 - k)p] \rho_1(m=1)}{[k + (1 - k)p] \rho_1(m=1) + [k + (1 - k)q] (1 - \rho_1(m=1))}.\]

To satisfy $\rho_2(m=1, s=1) \geq \frac{1}{2}(1 - v)$, the maximum allowable $r_B$ for the complementarity strategy is:
\[r_B^{\text{comp}} = \min\left\{1,\frac{\left(\frac{p+k(1-p)}{q+k(1-q)}\right)\left(\frac{1+v}{1-v}\right)\left(\frac{\rho_{0}}{1-\rho_{0}}\right)-k}{1-k}\right\}.\]
Note that $r_B^{\text{comp}}\geq 0$ is true when $\underline{\underline{\rho}}(k) \leq \rho_0 < \overline{\overline{\rho}}(k)$.

Let the sender's expected payoff with each respective strategy be:
\[\pi^{CB}_{\text{self}} = \rho_0 + (1-\rho_0) \cdot r_B^{\text{self}},\]
and 
\[\pi^{CB}_{\text{comp}} = \rho_0 p  + (1-\rho_0)  q \cdot r_B^{\text{comp}}.\]

The sender prefers self-sufficiency if it is feasible ($p\leq p_1$) and it yields higher profit, i.e., $\pi^{CB}_{\text{self}} \geq \pi^{CB}_{\text{comp}}$. Solving for $p$ gives the threshold value of $\bar{\bar{p}}$, given by:
%\[\hat{\rho}  = \frac{(v-1) (k (q-1)-q) ((k-1) q+1)}{(1-k) p ((k-1) q (v-1)-2)+(k-1)^2 q^2 (v-1)+2}, \]
\[\bar{\bar{p}} \equiv \min\{p_1,p_2\},\]
where 
\begin{align*}
p_1 & = \frac{\rho _0 (k (v-1) ((k-1) q+1)-v-1)-k (v-1) ((k-1) q+1)}{(k-1) \rho _0 (v+1)}, \\
p_2 &= \frac{\text{vRatio}(kq(-kq+q-2)+k+q)+(q+k(1-q))(1-\left(1-k\right)q)(k(q-\text{rRatio}-1)+\text{rRatio})}{(k-1)\left(-q+k^{2}(-1+q)q+q^{2}+k(-1-2(-1+q)q)+(-2+q)q+k(-1+(1-q)q)\text{vRatio}\right)}, 
\end{align*}

$\text{rRatio}=\frac{\rho_0}{1-\rho_0}$, and $\text{vRatio}=\frac{1+v}{1-v}$.

Taking derivatives of both components of the threshold, $p_1$ and $p_2$, w.r.t. $\rho_0$ gives
\begin{align*}
\frac{\partial p_1}{\partial\rho_0} &=\frac{k(1-v)((k-1)q+1)}{\rho_{0}^{2}(1-k)(v+1)}, \\
 \frac{\partial p_2}{\partial\rho_0} &=\frac{k(1-q)(1-v)(k(1-q)+q)(1-(1-k)q)}{\rho_{0}^{2}(1-k)\left(k^{2}(1-q)q(1-v)+k(2-(1-q)q(3-v))+q(3+v-2q)\right)},
\end{align*}
It is straightforward to see that both the numerators and denominators are positive in both expressions, implying that $\frac{\partial \bar{\bar{p}}}{\partial \rho_0} > 0$.
\qed

\paragraph{Proof of Proposition \ref{prop: CB-impact}}

Table \ref{tab:rB-cB} below summarizes the sender's optimal strategy ($r_B^*$) in the case of confirmation bias.

\begin{table}[h!]
\centering
\resizebox{\textwidth}{!}{%
\begin{tabular}{@{}p{5.5cm}p{5cm}p{8.5cm}@{}}
\toprule
\textbf{(Non-Automatic) \newline Parameter Region} & \textbf{Equilibrium Strategy} & \textbf{Value of $r_B^*$} \\ 
\midrule
$p \leq \bar{\bar{p}}$ or $\rho_0 \geq \hat{\rho}$ 
& Self-sufficiency & 
$\displaystyle \frac{\left(\frac{1-(1-k)p}{1-(1-k)q}\right)\left(\frac{1+v}{1-v}\right)\left(\frac{\rho_0}{1-\rho_0}\right)-k}{1-k}$ \\
\addlinespace[0.5em]
 $p > \bar{\bar{p}}$ and $\rho_0<\hat{\rho}$

& Complementarity & 
$\displaystyle \min\left\{1,\frac{\left(\frac{p+k(1-p)}{q+k(1-q)}\right)\left(\frac{1+v}{1-v}\right)\left(\frac{\rho_0}{1-\rho_0}\right)-k}{1-k}\right\}$ \\
\bottomrule
\end{tabular}%
}
\caption{\footnotesize{The equilibrium values of $r_B^*$ (under confirmation bias) across different parameter regions. Specifically, $\hat{\rho}  = \frac{(v-1) (k (q-1)-q) ((k-1) q+1)}{(1-k) p ((k-1) q (v-1)-2)+(k-1)^2 q^2 (v-1)+2}$ is derived by equating profits from the complementarity and self-sufficiency strategies when $r_B^{\text{self}} < r_B^{\text{comp}} = 1$.}}
\label{tab:rB-cB}
\end{table}

First, we prove that $\frac{\partial r_B^{\text{comp}}}{\partial k}\leq 0$, for any $\underline{\underline{\rho}}(k)\leq \rho_0 \leq \overline{\overline{\rho}}(k)$. 

Let $\Lambda(\rho_0) = \left(\frac{1+v}{1-v}\right)\left(\frac{\rho_{0}}{1-\rho_{0}}\right)$, $N(k) = p+k(1-p)$, and $D(k) = q+k(1-q)$. 

We can rewrite $r_B^{\text{comp}}$ as follows:
\[r_B^{\text{comp}} = \min\left\{1, R(k)\right\} \quad \text{where} \quad R(k) = \frac{\Lambda(\rho_0) \frac{N(k)}{D(k)} - k}{1-k}.\]
Note that imposing $R(k)\leq 1$ yields an upper bound for $\rho_0\leq\rho^*$, where  $\Lambda(\rho^*)=\frac{D(k)}{N(k)}$.

Taking derivatives of $R(k)$ w.r.t. $k$, we obtain:
\begin{equation}\label{eq: R}
\frac{\partial R}{\partial k}=\frac{\Lambda(\rho_{0})\cdot Z(k)-1}{\left(1-k\right)^{2}},
\end{equation}
where  
\[Z(k)=\frac{\left(q-p\right)\left(1-k\right)+N(k)D(k)}{D(k)^{2}}.\]
Note that $\Lambda(\rho_0)$ is positive and strictly increases in $\rho_0$. The sign of the derivative thus depends on the sign of $Z(k)$, since the denominator is strictly positive.

If $Z(k)\leq 0$, then it directly follows that the numerator of \eqref{eq: R} is negative. If $Z(k)>0$, then we must have 
\begin{align*}
\Lambda(\rho_0) Z(k)-1 & \leq \Lambda(\rho_0=\rho^*)\cdot Z(k)-1 \\
& = \frac{D(k)}{N(k)} \cdot\frac{\left(q-p\right)\left(1-k\right)+N(k)D(k)}{D(k)^{2}}-1 \\
& = \frac{\left(q-p\right)\left(1-k\right)}{D(k)N(k)} \\
&<0.
\end{align*}
The last inequality follows from $0<q<\frac{1}{2}<p<1$ and $0<k<1$. That is, the numerator is always negative under specified conditions. 
Hence we can conclude that $\frac{\partial r_B^{\text{comp}}}{\partial k}\leq 0$.

\medskip{}

Next, taking derivatives of $r_B^{\text{self}}$ w.r.t. $k$, we obtain:
\[\frac{\partial r_B^{\text{self}}}{\partial k} = \frac{\rho _0 [(k-1) q ((k-1) p (v+1)+q (k (-v)+k+v-1)+4)+2]+(v-1) ((k-1) q+1)^2}{\left(\rho _0-1\right) (v-1) \left((k-1)^2 q+k-1\right)^2}.\]
Note that the denominator is strictly positive, whereas the sign of the numerator monotonically depends on $\rho_0$. Moreover, we have 
\[\frac{\partial^2 r_B^{\text{self}}}{\partial k \partial \rho_0}=\frac{(v+1) ((k-1) q ((k-1) p+2)+1)}{(k-1)^2 \left(\rho _0-1\right){}^2 (1-v) ((k-1) q+1)^2}>0.\]
Hence we have $\frac{\partial r_B^{\text{self}}}{\partial k }>0$ iff $\rho_0>\rho^+$, given by 
\[\rho^+\equiv \frac{(v-1) ((k-1) q+1)^2}{(k-1) q (-(k-1) p (v+1)+(k-1) q (v-1)-4)-2}.\]
It is easy to verify that $\rho^{+}$ is strictly positive given that $q<\frac{1}{2}$.  

\qed

\paragraph{Proof of Proposition \ref{prop: multi}} 

Recall that $\alpha_M$ denotes the share of receivers who only observe the message, and $\alpha_{MS}$ those who observe both the message and the signal. Let $r_B^*$ denote the sender’s equilibrium strategy in the benchmark model, and $r_B^{MR}$ their strategy in the multi-receiver setting.

In the multi-receiver setting, the sender’s payoff is a weighted average of support from two groups: $M$ and $MS$. The former group’s posterior belief depends effectively on $\rho_1$, whereas the latter’s depends on $\rho_2$.

In the benchmark model, the sender chooses between two candidate strategies depending on how much validation is needed from the signal:

\begin{itemize}
\item A \textit{self-sufficiency} strategy ensures support under both signal realizations. Here, the sender is inauthentic with probability $r_B^{\text{self}}$ such that $\rho_2(m=1,s=0) \geq \frac{1}{2}(1-v)$.
\item A \textit{complementarity} strategy secures support only when $s=1$, where the sender is inauthentic with probability $r_B^{\text{comp}}$ such that $\rho_2(m=1,s=1) \geq \frac{1}{2}(1-v)$ but $\rho_2(m=1,s=0) < \frac{1}{2}(1-v)$.
\end{itemize}

These strategies remain optimal for persuading receivers in group $MS$, who observe both the message and the signal. However, the presence of a second receiver group ($M$) introduces an additional consideration: whether and how to persuade those who observe only the sender’s message.

Let $r_B^0$ denote the maximum probability of misrepresentation that still ensures support from the $M$-type receiver, i.e., $\rho_1(m=1) \geq \frac{1}{2}(1-v)$. This defines a third strategic option: a “direct persuasion” strategy.

The corresponding optimal values of $r_B$ under each strategy are:
\begin{align*}
& r_B^{\text{self}} = \frac{(1 - p)(1 + v) \rho_0}{(1 - q)(1 - v)(1 - \rho_0)} & \text{(self-sufficiency)}, \\
& r_B^{\text{comp}} = \frac{p(1 + v) \rho_0}{q(1 - v)(1 - \rho_0)} & \text{(complementarity)}, \\
& r_B^0 = \frac{(1 + v) \rho_0}{(1 - v)(1 - \rho_0)} & \text{(direct persuasion)}.
\end{align*}

It is straightforward to verify that $r_B^{\text{self}} < r_B^0 < r_B^{\text{comp}}$.

To characterize how the sender’s strategy adapts to receiver composition, we consider two major cases based on whether the prior $\rho_0$ is above or below the threshold $\overline{\rho}$ identified in Result 1.

\textbf{Case 1: $\rho_0 > \overline{\rho}$} \\
As shown in Result 1, the receiver automatically supports the sender regardless of the signal realization. As a result, the sender has no incentive to change their strategy across receiver types. The message alone is sufficient to induce support, and the sender continues to use the ``automatic affirmation'' strategy: $r_B^{MR} = r_B^* = 1$.

\textbf{Case 2: $\rho_0 < \overline{\rho}$} \

The informativeness of the message becomes relevant. The sender's strategy now depends on the relative share of receivers who observe the signal versus those who do not. Let $\underline{\alpha}$ denote the threshold in $\frac{\alpha_M}{\alpha_{MS}}$ above which switching strategies becomes profitable. We distinguish two subcases depending on whether the sender adopts self-sufficiency or complementarity strategy in the benchmark model:

\begin{itemize}
\item \textbf{Case 2.1: $p\leq \overline{p}$ or $\hat{\rho}<\rho_{0}<\overline{\rho}$}  \\
This is the region where the benchmark strategy is self-sufficiency: $r_B^* = r_B^{\text{self}}$. Since $r_B^{\text{self}}<r_B^0$, staying with self-sufficiency gains support from both $M$ and $MS$ type receivers. Meanwhile, increasing $r_B$ to $r_B^0$ increases the frequency of prosocial message while maintaining support from group $M$ receivers, but decreases its information value, potentially causing a loss of support from group $MS$ receivers when $s=0$.

The sender thus compares profits  from switching to the direct persuasion strategy for group $M$, given by
\[\pi^{MR}_{\text{direct}}=\alpha_M(\rho_0 + (1 - \rho_0) r_B^0)+\alpha_{MS}(\rho_0 p +(1-\rho_0)r_B^0 q),\]
with profit from retaining self-sufficiency, given by 
\[\pi^{MR}_{\text{self}}=(\alpha_M + \alpha_{MS})(\rho_0 + (1 - \rho_0) r_B^{\text{self}}).\]
Switching to more frequent lying is optimal iff $\pi^{\text{direct}}>\pi^{\text{self}}$, or equivalently, 
\[\frac{\alpha_M}{\alpha_{MS}} >\frac{(1-v)(1-p)(1-q)+(1+v)(1-p-q+q^2)}{(1+v)(p-q)}.\]

\medskip{}

\item \textbf {Case 2.2: $p>\overline{p}$ and $\rho_{0}<\hat{\rho}$} \\
This is the region where the benchmark strategy is complementarity: $r_B^* = r_B^{\text{comp}}$. Since $r_B^{\text{comp}}>r_B^0>r_B^{\text{self}}$, staying with complementarity strategy only secures support from $MS$ receivers who see a positive signal, while reducing $r_B$ (to $r_B^0$ or $r_B^{\text{self}}$) makes the message more persuasive, potentially gaining support from $M$ and even $MS$ receivers who see a negative signal, but reduces the frequency of prosocial message.

Now the sender compares the payoffs across three different strategies:
Staying with complementarity:
\[\pi^{MR}_{\text{comp}} = \alpha_{MS}(\rho_0 p + (1 - \rho_0) r_B^{\text{comp}} q)\]
Switching to direct persuasion for group $MS$ observing $s=1$ and group $M$:
\[\pi^{MR}_{\text{direct}} = \alpha_M(\rho_0 + (1 - \rho_0) r_B^0)  + \alpha_{MS}(\rho_0 p + (1 - \rho_0) r_B^0 q)\]
Or use self-sufficiency to persuade both groups (regardless of $s$):
\[\pi^{MR}_{\text{self}} = (\alpha_M+\alpha_{MS})(\rho_0 + (1 - \rho_0) r_B^{\text{self}})\]
The sender stays with complementarity $r_B^{MR} = r_B^*=r_B^{\text{comp}}$ iff $\pi^{MR}_{\text{comp}}\geq \max\{\pi^{MR}_{\text{self}},\pi^{MR}_{\text{direct}}\}$.
In other words, switching to less frequent lying (either by adopting direct persuasion or self-sufficiency) is optimal if and only if 
\[\frac{\alpha_M}{\alpha_{MS}} > \min\left\{
 \frac{1}{2}(1+v)(p-q),\frac{2p(1-q)}{2-p\left(1+v\right)-q\left(1-v\right)}-1\right\}.\]
\end{itemize}

\qed